\documentclass[fleqn,usenatbib]{mnras}

\usepackage{newtxtext,newtxmath}
\usepackage[T1]{fontenc}

\makeatletter
\DeclareRobustCommand{\VAN}[3]{#2}
\let\VANthebibliography\thebibliography
\def\thebibliography{\DeclareRobustCommand{\VAN}[3]{##3}\VANthebibliography}

\usepackage{natbib}

\usepackage{graphicx}
\usepackage{stfloats}
\usepackage{textcomp}
\usepackage{float}
\usepackage{subfig}
\usepackage{starfont}
\usepackage{amsmath}

\usepackage{lineno}

\newcommand{\fotwo}{$f$O$_2$}
\newcommand{\orcid}[1]{\href{https://orcid.org/#1}{\includegraphics[height=1.5ex]{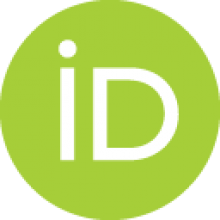}}}

\title[Characterization of redox states with LIFE]{Characterizing the oxidation state of rocky exoplanets with the Large Interferometer for Exoplanets (LIFE)}
\author[L. Cesario et al.]{Lorenzo Cesario \orcid{0000-0002-3286-7683},$^{1}$\thanks{Email: lorenzo.cesario.research@gmail.com}
        Tim Lichtenberg \orcid{0000-0002-3286-7683},$^{1}$\thanks{Email: tim.lichtenberg@rug.nl} 
        Michiel Min \orcid{0000-0001-5778-0376},$^{2}$
        Lena Noack \orcid{0000-0001-8817-1653},$^{3}$ 
        Caroline Brachmann \orcid{0009-0006-4753-7536},$^{3}$ 
        \newauthor Eleonora Alei \orcid{0000-0002-0006-1175},$^{4}$ 
        Sascha P. Quanz \orcid{0000-0002-0006-1175},$^{5,6}$ 
        and the LIFE Collaboration$^{7}$ \\
\\
$^{1}$Kapteyn Astronomical Institute, University of Groningen, The Netherlands \\
$^{2}$SRON Netherlands Institute for Space Research, Leiden, The Netherlands \\
$^{3}$Freie Universität Berlin, Institute of Geological Sciences, Berlin, Germany \\
$^{4}$NPP Fellow, NASA Goddard Space Flight Center, Greenbelt, MD, USA \\
$^{5}$ETH Zurich, Institute for Particle Physics \& Astrophysics, Wolfgang-Pauli-Str. 27, 8093 Zurich, Switzerland\\
$^{6}$ETH Zurich, Department of Earth and Planetary Sciences, Sonneggstrasse 5, 8092 Zurich, Switzerland\\
$^{7}$\href{https://www.LIFE-space-mission.com}{LIFE-space-mission.com}
}

\date{Accepted XXX. Received YYY; in original form ZZZ}

\pubyear{\the\year{}}

\begin{document}

\label{firstpage}
\pagerange{\pageref{firstpage}--\pageref{lastpage}}
\maketitle

\begin{abstract}

\noindent The oxidation state of rocky exoplanets is expected to play a fundamental role in shaping the chemical composition of their secondary atmospheres by influencing the chemical composition of volcanically released gasses. Distinguishing planetary redox states through direct atmospheric characterization would offer insight into the formation and evolution of secondary atmospheres on exoplanets and inform the background chemistry of putative biosignatures. The Large Interferometer For Exoplanets (LIFE) mission concept aims to employ a space-based mid-infrared nulling interferometer to characterize exoplanetary atmospheres. In this work, we assess LIFE’s performance in distinguishing the redox states of rocky exoplanets by direct spectroscopic measurements. We focus on the observability and spectral features of redox-sensitive molecules in secondary atmospheres of Earth-sized exoplanets. We develop and apply a retrieval framework based on the ARtful modeling Code for exoplanet Science (ARCiS) and the LIFE mission simulator (LIFEsim) to simulate observations of Earth-sized planets with atmospheres from a range of plausible mantle redox conditions. Our simulations show that LIFE in its baseline configuration can successfully constrain dominant atmospheric species (e.g. CO\textsubscript{2}, CH\textsubscript{4} and NH\textsubscript{3}) with sufficient accuracy to distinguish redox states for planets orbiting a Sun-like star at 10 pc. Retrieved redox-sensitive molecules show clear trends across oxidation states, with CO\textsubscript{2} dominating in oxidizing (with oxygen fugacity \fotwo $\sim$ IW+2 to IW+6, where IW is the iron-w\"ustite buffer) environments and NH\textsubscript{3} in reducing (\fotwo $\sim$ IW-2 to IW-6) environments, and CH\textsubscript{4} serving as a strong tracer among intermediate (\fotwo $\sim$ IW+4 to IW-4) oxidation states.

\end{abstract}


   \maketitle
%

\section{Introduction}

The Large Interferometer For Exoplanets (LIFE) collaboration is developing a space mission with the goal of detecting and characterizing terrestrial exoplanets and to search for biosignatures \citep{quanz_et_al._2022,glauser2024}. The mission concept consists of a space-based nulling interferometer with (in its baseline architecture) four collecting telescopes in an X-array configuration, observing in the mid-infrared region of the spectrum. Previous simulations of the instrument's performance show promising results for the detection of temperate, potentially habitable exoplanets \citep{dannert_et_al._2022,2022A&A...664A..23K,2023A&A...673A..94K,2022A&A...665A.106A,2022A&A...668A..52K,2023AsBio..23..183A,2024AJ....167..128A,2023A&A...678A..96C,Janson2023AA} as well as for the detection of hot, transient, magma oceans in early stages of a planet's evolution track \citep{2024A&A...692A.172C,bonati2019}. A considerable amount of work has already been carried out studying the instrument designs and their resulting performances, source contamination, and accessible parameter space \citep{2022A&A...664A..52H,2023A&A...670A..57H,dannert_et_al._2022, quanz_et_al._2022, 2025AJ....169..244C,glauser2024}. 

In this work, we focus on the potential of the LIFE mission to distinguish the redox states of temperate terrestrial exoplanets. The redox state of a planet is an important factor in planetary science, playing a major role in the evolution and composition of the secondary atmosphere of terrestrial planets \citep{lichtenberg2023,lichtenberg2025}. The redox state of an environment is a measure of its chemical tendency to gain (reduction) or lose electrons (oxidation). In other words, it describes how reducing or oxidizing the environment is, quantitatively measured by the activity of oxygen in the chemical environment, and it plays a crucial role in volcanic outgassing, governing the gases that are released in the atmosphere \citep{gaillard2021,suer2023}.The redox state of a planetary mantle yields insights into which molecules dominate the atmosphere, and vice versa. Distinguishing the redox state of an abiotic planet by direct observation of its secondary atmosphere can provide us with important information about the  background atmospheric environments, giving us insight into the production of secondary atmospheres under different compositions and thermal histories \citep{wordsworth2022,nicholls2024,nicholls25c}. This will serve as a necessary stepping stone to be able to decide between the potential biogenicity of anomalous atmospheric signals from a planet. In order to quantify chemical disequilibrium we need to understand the possible landscape of abiotic environments of rocky exoplanets \citep{2022NatAs...6..189K,Lichtenberg2025Science}. The redox state of terrestrial exoplanets may substantially differ from modern Earth, influenced by varying core-mantle ratios, different volatile abundances, different ratios of refractory species, and different geodynamic modes \citep{Guimond2023MNRAS,Guimond2024RvMG,lichtenberg2025,Baumeister2025SSRv}, motivating us here to probe a widely varying array of oxidation states.

In this work we will focus on the long-term steady-state evolution of the secondary atmospheres of rocky exoplanets, i.e., we do not compute self-consistent planetary evolution sequences, but we assume that volcanic degassing, as operating on Earth today, will set the chemical composition of the secondary, abiotic atmosphere \citep{Ortenzi2020NatSR,Guimond2021PEPI,2020E&PSL.55016546L,2022JGRE..12707123L, liggins2022, liggins2023, 2022PSJ.....3...93B, Drant2025, gkouvelis2025}. We will thus work with key molecules which are anticipated to form the dominant background gases for secondary atmospheres on rocky and terrestrial exoplanets \citep{gaillard2021,lichtenberg2025}. \citet{2025Icar..42916450B} developed a three-step model, coupling mantle and atmospheric composition, which simulates the production of secondary atmospheres on a planet. Most importantly for this study, their model computes the resulting partial pressures of six major molecules in the planet's secondary atmosphere, namely: H\textsubscript{2}, H\textsubscript{2}O, CO\textsubscript{2}, NH\textsubscript{3}, CH\textsubscript{4} and N\textsubscript{2}. It should be mentioned here that this accounts for cold atmospheres ($\leq 600$K), whereas in hotter atmospheres CO and sulfur gases would occur as well. Here, we explicitly exclude the possibility of life polluting this environment, and we ask the basic question: if there is no life, can LIFE distinguish the redox conditions of a particular exoplanet? This study thus builds the stepping stone toward more detailed analyses of how a LIFE-like mission may characterize the atmosphere of terrestrial exoplanets, to arrive at the mission requirements to distinguish biogenic from abiotic signals.

\section{Methods}
\hypertarget{methods}{}
\label{sec: Methods}

In this work, we use the ARtful modelling Code for exoplanet Science (ARCiS; \citealt{2020A&A...642A..28M, 2022A&A...665A...2C}) as our main framework for generating planetary emission spectra and retrieving atmospheric composition. Our main developed procedure is as follows: we take planetary characteristics representative of varying oxidation states and thermodynamic conditions from \cite{2025Icar..42916450B} and generate the emission flux and pressure–temperature structure with the ARCiS forward model (assuming Earth-like values for other parameters). We then simulate observations with the LIFE mission simulator LIFEsim \citep{dannert_et_al._2022} to obtain astrophysical noise. This gives us synthetic LIFE observations of Earth-sized planets orbiting Sun-like stars at 1 AU and placed at 10 pc. We then run a Bayesian retrieval with ARCiS to constrain the simulated planets’ characteristics and molecular partial pressures, based on the assumptions used in both the modeling and retrieval routines. At the end of the process, we perform a statistical test, following \citet{2022A&A...665A.106A}, to evaluate how well LIFE observations can distinguish between different planetary redox states. In the following we describe in more detail the individual steps of this routine.

\subsection{Secondary Atmosphere Modeling}

\cite{2025Icar..42916450B} developed a coupled mantle-atmosphere model to simulate the formation and long-term evolution of secondary atmospheres on rocky exoplanets. The model integrates mantle melting, melt ascent, and volcanic degassing processes in order to link the planet’s interior redox state, melt production rates, and volatile content (including volatile partitioning between mantle and melt) to the resulting atmospheric properties. Oxygen fugacity rates range from $\Delta$ IW-6 to $\Delta$ IW+6 relative to the Iron Wustite buffer. The average melt production rate adopted in this study is $10^{15} kg  yr^{-1}$, corresponding to a scenario of 10\% intrusive-to-extrusive ratio, following the calculation by \citet{Guimond2021PEPI}. However, the melt production rates are varied between $10^{13} kg  yr^{-1}$ and $10^{15} kg  yr^{-1}$ to test for the influence of different melting fluxes. In particular, we explore cases with higher melt production rates to test the retrieval framework on cases with high H\textsubscript{2} abundances. The atmospheric module simulates equilibrium gas chemistry utilizing the FastChem code, including water condensation and energy-limited hydrogen escape. By varying parameters such as mantle oxygen fugacity, surface temperature, and melt production rates, the model explores how these affect the resulting atmospheric composition (via the resulting mixing ratios of six major molecules, i.e. H\textsubscript{2}, H\textsubscript{2}O, CO\textsubscript{2}, NH\textsubscript{3}, CH\textsubscript{4} and N\textsubscript{2}) and surface pressure. The initial volatile budget in the simulations used in \cite{2025Icar..42916450B} is the one estimated for Earth after magma ocean solidification, namely 10 ppm N\textsubscript{2} set by \citet{2025Icar..42916450B}, and 450 ppm H\textsubscript{2}O and 50 ppm CO\textsubscript{2} from \citet{elkins-tanton2008}. These values, together with the redox state and imposed melt surface flux determine the initial volatile content. Furthermore, the surface temperature is set as initial condition and varied across simulations rather than being dinamically calculated. This is done to explore its effect on atmospheric composition in \cite{2025Icar..42916450B}. The degassing model of \citet{2025Icar..42916450B} models the secondary atmosphere with a prescribed surface temperature and surface pressure, without resolving a vertical (1D) structure or imposing layer-by-layer chemical equilibrium. A vertically resolved 1D atmosphere, together with a self-consistent radiative–convective temperature profile, is computed only later in this work by ARCiS. The main steps of the model are as follows: firstly, the melt is produced in the mantle and its volatile content is set by the ambient parameters (i.e. redox state, pressure, temperature, degree of melting, initial volatile content). Then, the speciation code calculates the degassing of volatiles using solubility laws for the exsolution of volatiles and thermodynamical equilibrium for gas speciation. A table containing the references for solubility and partitioning coefficients used in \cite{2025Icar..42916450B} can be found in Table \ref{tab:solubility}. Finally, the model calculates water condensation, redistribution (through FastChem) and hydrogen escape. The model then determines the speciation of gases in the atmosphere taking the feedback effects of the existing atmosphere on the degassing into account. This is done by updating each species' atmospheric partial pressure after each degassing step and feeding these back to the solubility calculation of the next step. This way the revised partial pressure of each gas will affect its solubility, and consequently its degassing, in subsequent steps. The outputs provide atmospheric pressures and compositions of secondary atmospheres across a range of surface temperatures and mantle redox states, enabling us to observationally explore the mantle's redox state role in shaping the composition of exoplanetary atmospheres.

\subsection{ARCiS: modeling and retrieval}

The ARtful modelling Code for exoplanet Science (ARCiS) is a modeling framework for the analysis of exoplanet transmission and emission spectra \citep{2020A&A...642A..28M}. ARCiS makes use of modules for radiative transfer, equilibrium chemistry via GGchem, and self-consistent cloud formation, while allowing parameters such as molecular abundances, cloud properties, and temperature profiles to be freely retrieved. Its ultimate aim is to link observations of exoplanets to physical models of exoplanet atmospheres. Retrieval methods are essential to obtain atmospheric and planetary parameters from observed spectra with almost no prior information. ARCiS can employ a Bayesian retrieval approach that treats atmospheric properties, such as molecular partial pressures and planetary characteristics, as free, independent parameters with minimal physicochemical constraints. This enables exploration of a wide parameter space to fit observational data without enforcing strict chemical equilibrium or radiative convective $P$-$T$ profiles, adjusting to the limited prior knowledge of exoplanet atmospheres in observations. For parameter space exploration, ARCiS employs nested sampling methods  \citep[MultiNest algorithm, ][]{2008MNRAS.384..449F}, providing final posterior probability distributions for the parameters based on likelihood estimations.
The ARCiS framework has already been shown to be fast and flexible enough for retrievals of mid-infrared spectra with mission data such as JWST \citep{Dyrek2024Natur, Barrado2023Natur, Crouzet2025arXiv}. It also offers the possibility to generate, via its own radiative transfer model, the emission flux of a model planet. This allows us to maintain consistency between the radiative transfer model used to produce the planet's emission and the theoretical spectra in the retrievals. This avoids discrepancies in the fluxes due to distinct handling of physical and/or chemical processes, or due to the assumed opacity tables for major molecules, problems previously highlighted by \cite{AleiSPIE}. By using ARCiS in both processes we minimize these systematic effects that usually bring about discrepancies, and thus ensure a consistent interpretation of observational data. This coherence between forward modeling and retrieval ensures more reliable constraints on atmospheric properties and molecular abundances.

\subsection{Input spectra}

We use the atmospheric datasets generated by \citet{2025Icar..42916450B}, specifically the vertically constant volume mixing ratios of the six major species, together with the surface pressure. These are used as the input compositions from which we generate our forward-model emission spectra with ARCiS, on which we then perform the retrievals. Given the atmospheric data at hand, we use ARCiS to generate the atmospheric $P$-$T$ structure and spectra of planets with 600K surface temperature at different redox states. We choose this value of surface temperature, from the available \cite{2025Icar..42916450B} data, in order to study the underlying geochemical environment of hot rocky planets as a probe for earlier stages of temperate exoplanets. These scenarios provide us with insight into the formation and evolution processes of temperate exoplanets' secondary atmospheres. In Table \ref{tab:radiativetransfer} we provide the parameters we use to run ARCiS' radiative transfer model for six planets with different redox states quantified according to \fotwo. We indicate the oxidation state by deviations from the iron-wüstite (IW) buffer (from most reducing to most oxidising) in log units: IW-6.0, -3.8, -1.6, 1.6, 3.8 and 6.0. The buffer is a standard reference scale used to quantify the oxidating (or reducing) state of an environment, from oxidising (positive IW) to very reducing conditions (negative IW). For simplicity, we round these values from \citet{2025Icar..42916450B} to the nearest integer, i.e., IW-6, -4, -2, 2, 4 and 6 throughout the rest of the paper. For plots of the planet-specific volume mixing ratios and surface pressures we probe through in this work, we refer the reader to Fig.~\ref{fig:noack1} and ~\ref{fig:noack2}. We quantify the presence or absence of the following six molecules, H\textsubscript{2}, H\textsubscript{2}O, CO\textsubscript{2}, CH\textsubscript{4}, NH\textsubscript{3}, and N\textsubscript{2}, and assume they dominate the chemical environment in the secondary atmospheres of exoplanets in the relevant wavelength range. The resulting emission spectra can then be found in Fig.~\ref{fig:emission-retrievals} (red lines). Importantly, we do not modify the atmospheric composition according to the new $P$-$T$ profile (effectively de-coupling climate calculations with degassing and chemistry) and assume the atmospheric composition to be held fixed along pressure. As a consequence of this, the surface temperature of our forward model then differs from the initially imposed 600 K.

\subsection{Retrieval setup}

As a next step, we perform atmospheric retrievals on the simulated emission spectra using the ARCiS framework, treating each model parameter as an independent free parameter. This approach allows the retrieval to have maximum flexibility in exploring parameter space without enforcing chemical or radiative-convective equilibrium constraints. The retrieval includes key planetary properties such as planetary radius and mass, as well as the partial pressures of six molecular species in its atmosphere, namely H\textsubscript{2}, H\textsubscript{2}O, CO\textsubscript{2}, CH\textsubscript{4}, NH\textsubscript{3}, and N\textsubscript{2}. These are all sampled with flat log priors to reflect our minimal prior knowledge of the planet. The up to date line lists used for opacity computations in ARCiS can be found in Table \ref{linelist}. Moreover, collision-induced absorption (CIA) is accounted for, ensuring that key opacity contributions, other than  H\textsubscript{2}-H\textsubscript{2} and H\textsubscript{2}-He interactions, are also included in the forward model and retrieval. The list of CIA considered and their references can be found in Table \ref{tab: ciareferences}, although these lines' measurements do not cover the higher temperatures some of our targets reach. The full forward model set-up in ARCiS can be found in Table \ref{tab:radiativetransfer}. All retrievals are run with 300 MultiNest live points and the complete set of retrieval parameters and prior ranges is provided in Table \ref{tab: retrieval}.

\renewcommand{\arraystretch}{1.5}
\begin{table}
\centering
\caption{Parameter space used in the retrievals.}
\begin{tabular}{l|l|l}
\hline
\textbf{Parameter}              & \textbf{Range}                        & \textbf{Prior} \\ \hline
Planet Radius                   & 0.1 - 2.2 [R$_{\oplus}$]  & Flat linear    \\ \hline
Planet Mass                     & 0.3 - 1.9 [M$_{\oplus}$]  & Flat linear    \\ \hline
Temperature Points              &                                       &                \\ \hline
\textbf{Partial Pressures}       &                                       &                \\ \hline
log$_{10}$(H$_2$)               & $\mathcal{U}(-9,3)$                   & Flat log       \\ \hline
log$_{10}$(H$_2$O)              & $\mathcal{U}(-9,3)$                   & Flat log       \\ \hline
log$_{10}$(CO$_2$)              & $\mathcal{U}(-9,3)$                   & Flat log       \\ \hline
log$_{10}$(CH$_4$)              & $\mathcal{U}(-9,3)$                   & Flat log       \\ \hline
log$_{10}$(NH$_3$)              & $\mathcal{U}(-9,3)$                   & Flat log       \\ \hline
log$_{10}$(N$_2$)               & $\mathcal{U}(-9,3)$                   & Flat log       \\ \hline
\end{tabular}
\label{tab: retrieval}
\end{table}

\begin{table}
\centering
\caption{References for the line lists used in ARCiS' gas opacity computations.}
\begin{tabular}{lll}
\cline{1-2}
\textbf{Molecule} & \textbf{Reference}                       &  \\ \cline{1-2}
H$_2$             & HITRAN; \cite{GORDON20173}   &  \\ \cline{1-2}
H$_2$O            & ExoMol;  \cite{POLYANSKY2018} &  \\ \cline{1-2}
CO$_2$            & ExoMol; \cite{yurchenko_co2}     &  \\ \cline{1-2}
CH$_4$            & ExoMol;  \cite{YURCHENKO2017}   &  \\ \cline{1-2}
NH$_3$            & ExoMol;  \cite{COLES2019}       &  \\ \cline{1-2}
N$_2$             & ExoMol;  \cite{WESTERN2018127}    &  \\ \cline{1-2}
\end{tabular}
\label{linelist}
\end{table}

\begin{table}
\centering
\caption{References for the Collisionally Induced Absorption (CIA) data used in ARCiS}
\begin{tabular}{l|l|l}
\textbf{Pair}    & \textbf{Temperature{[}K{]}} & \textbf{ Reference}                                                                              \\ \hline
CO2-CO2 & 200 - 800                 & \cite{GRUSZKA1997172}     \\
&& \cite{BARANOV1999319} \\ && \cite{BARANOV201811} \\ \hline
N2-N2   & 70 - 400                  & \cite{chistikov2019} \\ && \cite{BARANOV2005160} \\
&& \cite{Lafferty:96}\\ && \cite{Hartmann} \\ \hline
N2-H2O  & 250 - 350                 & \cite{hartmann2018}                                                                  \\ \hline
CO2-CH4 & 200 - 350                 & \cite{wordworth2017}                                                                 \\ \hline
CO2-H2  & 200 - 350                 & \cite{wordworth2017}                                                                \\ \hline
H2-CH4  & 40 - 400                  & \cite{1986ApJ...304..849B}  \\ \hline                                                            
\end{tabular}
\label{tab: ciareferences}
\end{table}

The Bayesian retrieval approach explores the parameter space using the MultiNest nested-sampling algorithm \citep{2008MNRAS.384..449F}. Nested sampling evolves a fixed set of “live” points drawn from the prior towards regions of higher likelihood, progressively shrinking the enclosed prior volume and computing the Bayesian evidence. Forward models are used to generate spectra, and these model outputs are then compared to the observed (or simulated) fluxes. The posterior distribution of each parameter is obtained as a by-product of this process, from the likelihood-weighted set of sampled points. Thus inferring the most likely values for each parameter given the observed data. To ensure our retrieved temperature profile does not suffer from assumptions on the shape of the temperature structure (e.g. isothermal, adiabatic or parameterized) we retrieve a free temperature structure. We thus add six temperature points, along the pressure range, to the retrieved parameters. This increases model complexity but also ensures an agnostic approach to the atmospheric dynamics. Furthermore, the surface pressure is reconstructed as the sum of the retrieved partial pressures of the six major species. Since each molecule's partial pressure is anchored to its spectral features in the photospheric region that dominates the thermal flux emission, the summed surface pressure is observationally constrained. This follows the partial pressure parametrisation increasingly adopted in rocky-planet retrievals \citep[see e.g.,][]{hall2023, salvador2024, damiano_effects_2025} in which the gas partial pressures are treated as free, independent parameters and the surface pressure is defined as their sum.

\subsection{Instrument setup}

To simulate the performance of the LIFE mission concept, we use the LIFE mission simulator \citep[LIFEsim,][]{dannert_et_al._2022}. LIFEsim includes the contribution of the major astrophysical noises, such as photon noise and dust emission from the local and exozodiacal disks. The dust emission model in LIFEsim assumes that its distribution is similar to the Solar System's and for this work we set the dust in the observed system at three times the solar system level (i.e. a zodi of z=3). The disks are also assumed to always be face-on, symmetric and homogeneous. A study of the impact of angled disks on observations with LIFEsim can be found in \citet{quanz_et_al._2022}. In Table \ref{tab:instrument_values} we provide all the relevant simulation parameters.

In this study, the LIFEsim tool is used as a noise simulator. All the modelled planets are observed with the LIFE instrument until the simulation reaches an overall signal-to-noise ratio (S/N) of 10. This is calculated using the following equation:

\begin{equation}
    (S/N)_{tot} = \sqrt{\sum_{\lambda}{\left(\frac{S}{N}\right)^2_{\lambda}}}
\label{signall}
\end{equation}

\noindent taken from \cite{dannert_et_al._2022}. Once this value is reached we stop the observation and extrapolate the noise from the resulting S/N per wavelength bin. We obtain the noise distribution from our observation by dividing the planet's flux by the S/N to obtain a $\sigma$ factor. We then generate, at every wavelength bin, a white Gaussian noise (WGN) realization with zero mean and a standard deviation equal to the $\sigma$ factor. This approach produces a random noise instance at every wavelength bin which fluctuates around the true signal. This is done to reflect the behavior of photon noise in real observational data, to represent what an instrument would actually measure, rather than having a fixed $\sigma $ envelope. Essentially, we aim to obtain what an actual observation with LIFE would look like given the expected S/N by modeling the effect of the uncertainty on our spectrum rather than how uncertain we are of the measurement. We sample the noise instead of assuming it. This is an upgrade to the previously used methods where the noise was treated as an envelope on the spectrum, as it reflects the stochastic nature of observational noise. We note that because the mock observations are generated by scattering the model spectrum according to the predicted observational uncertainties, the retrieved posterior medians are not expected to match the exact input parameter values. Instead, they are expected to fall within approximately the 1$\sigma$ credible interval around the true values. Furthermore, we also note that the wavelength range considered in this study is restricted to 6-16 microns, based on the most recent LIFE mission instrument updates. This is a more conservative estimate with respect to recent LIFE retrieval studies with 4-18.5 microns \citep{2025MNRAS.tmp.1848G}. Thus, we expect to find looser constraints on molecules such as H\textsubscript{2}O with key spectral features wavelength bins lost to this restriction.

\begin{table}
\center

\caption{LIFEsim instrument setup and major parameters. See \citet{dannert_et_al._2022} for the definition and detailed description of the parameters.}
\label{longlist}
\begin{tabular}{lll}
\hline
\textbf{PARAMETER} & \textbf{UNIT}         & \textbf{DEFAULT}     \\ \hline
\begin{tabular}[c]{@{}l@{}}Aperture diameter \end{tabular}                                     & {[}m{]}      & 2           \\
\begin{tabular}[c]{@{}l@{}}Minimum wavelength\end{tabular}                                    & {[}\text{\textmu}m{]} & 6           \\
\begin{tabular}[c]{@{}l@{}}Maximum wavelength\end{tabular}                                    & {[}\text{\textmu}m{]} & 16        \\
\begin{tabular}[c]{@{}l@{}}Quantum efficiency\end{tabular}                                    & {[}\%{]}     & 70          \\
\begin{tabular}[c]{@{}l@{}}Photon throughput \end{tabular}                                    & [\%]         & 5            \\
\begin{tabular}[c]{@{}l@{}}Spectral resolution\end{tabular}                                   &             & 100          \\
\begin{tabular}[c]{@{}l@{}}Minimum nulling length\end{tabular}                              & {[}m{]}      & 10          \\
\begin{tabular}[c]{@{}l@{}}Maximum nulling length\end{tabular}                              & {[}m{]}      & 100         \\
\begin{tabular}[c]{@{}l@{}}Imaging/Nulling length ratio\end{tabular}                        &       & 6           \\
\begin{tabular}[c]{@{}l@{}}Slewing time\end{tabular}                                          & {[}s{]}      & 36000         \\
\begin{tabular}[c]{@{}l@{}}Time efficiency\end{tabular}                                       & {[}\%{]}     & 80          \\ \hline
\begin{tabular}[c]{@{}l@{}}Image size\end{tabular} &                                      {[}pixel{]}           & 256         \\
\begin{tabular}[c]{@{}l@{}}Optimal wavelength \end{tabular}                                   & {[}\text{\textmu}m{]} & 15          \\ \hline
\begin{tabular}[c]{@{}l@{}}Local zodi model\end{tabular}                                       & -            & "darwinsim" \\
\begin{tabular}[c]{@{}l@{}}Habitable zone model\end{tabular}                                  & -            & "MS"        \\ \hline
\\
\end{tabular}
\label{tab:instrument_values}
\end{table}

\subsection{Statistical test}

In order to determine whether we can distinguish the redox states of the observed planets, we calculate a quantitative separation between the retrieved cumulative posterior distributions of two redox states, for each molecule. This approach is taken from \cite{2022A&A...665A.106A} and is based on the Kolmogorov-Smirnov test, which assesses whether two samples are drawn from the same distribution (i.e., from the same underlying redox state) \citep{kolmogorov}. As per \cite{2022A&A...665A.106A}, we calculate the cumulative distributions $G^M(z)$ at every point $z$ over our prior range [$x_{min},x_{max}$] = [$10^{-9},10^3$]:
\begin{equation}
    G^M(z) = \frac{1}{N}\sum_{i=1}^N \textbf{P}_{x_i\leq z}
\end{equation}
where \textbf{P} is the posterior function, adding up the probabilities of all bins for $x_i\leq z$ and N is the total number of counts (used as normalization factor).
We then compare the resulting cumulative distributions of two different redox states (e.g. \textit{a} and \textit{b}) (for each molecule separately) by simply considering the maximum difference $\Delta$ between them, defined as:
\begin{equation}
    \Delta = \text{max }|G^M_{IW_a}(z) - G^M_{IW_b}(z)|
\end{equation}

The $\Delta$ $\in$ [0,1] factor will be used as a quantitative measure of how likely the considered model parameter's posterior distributions originate from two different underlying redox states. The idea is that larger values of $\Delta$ imply the two posterior distributions being considered are likely to originate from a different underlying redox state while lower values of $\Delta$ imply that we cannot confidently differentiate between the underlying redox states of the distribution being analysed. A $\Delta = 1$ indicates that there is a high likelihood the underlying redox state of the compared distributions is different.

\section{Results}

\begin{figure*}
    \centering
    \includegraphics[width=0.9\textwidth]{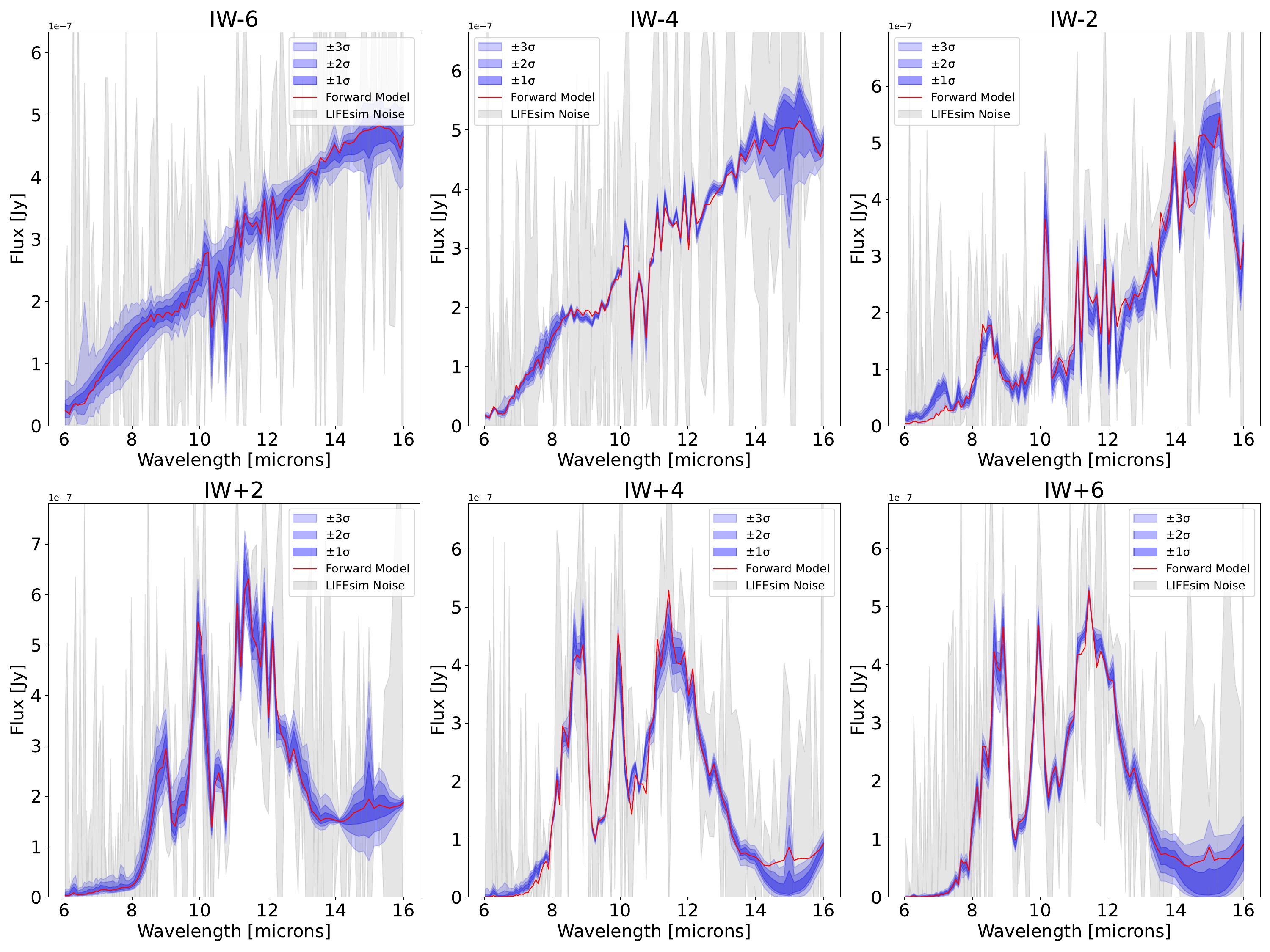}
    \caption{Retrieved emission spectra (blue regions) plotted with the true planet's flux from our forward model (red) for each redox state considered (IW-6 to IW+6). The retrieved spectra denser region represents the 1$\sigma$ confidence level, intermediate region 2$\sigma$, thin region the 3$\sigma$ level. The observational noise from LIFEsim is shown as a gray envelope. The retrieved spectra capture most spectral features in observed planets, especially in the more pronounced oxidised cases (bottom row).}
    \label{fig:emission-retrievals}
\end{figure*}

\begin{figure*}
    \centering
    \includegraphics[width=0.9\textwidth]{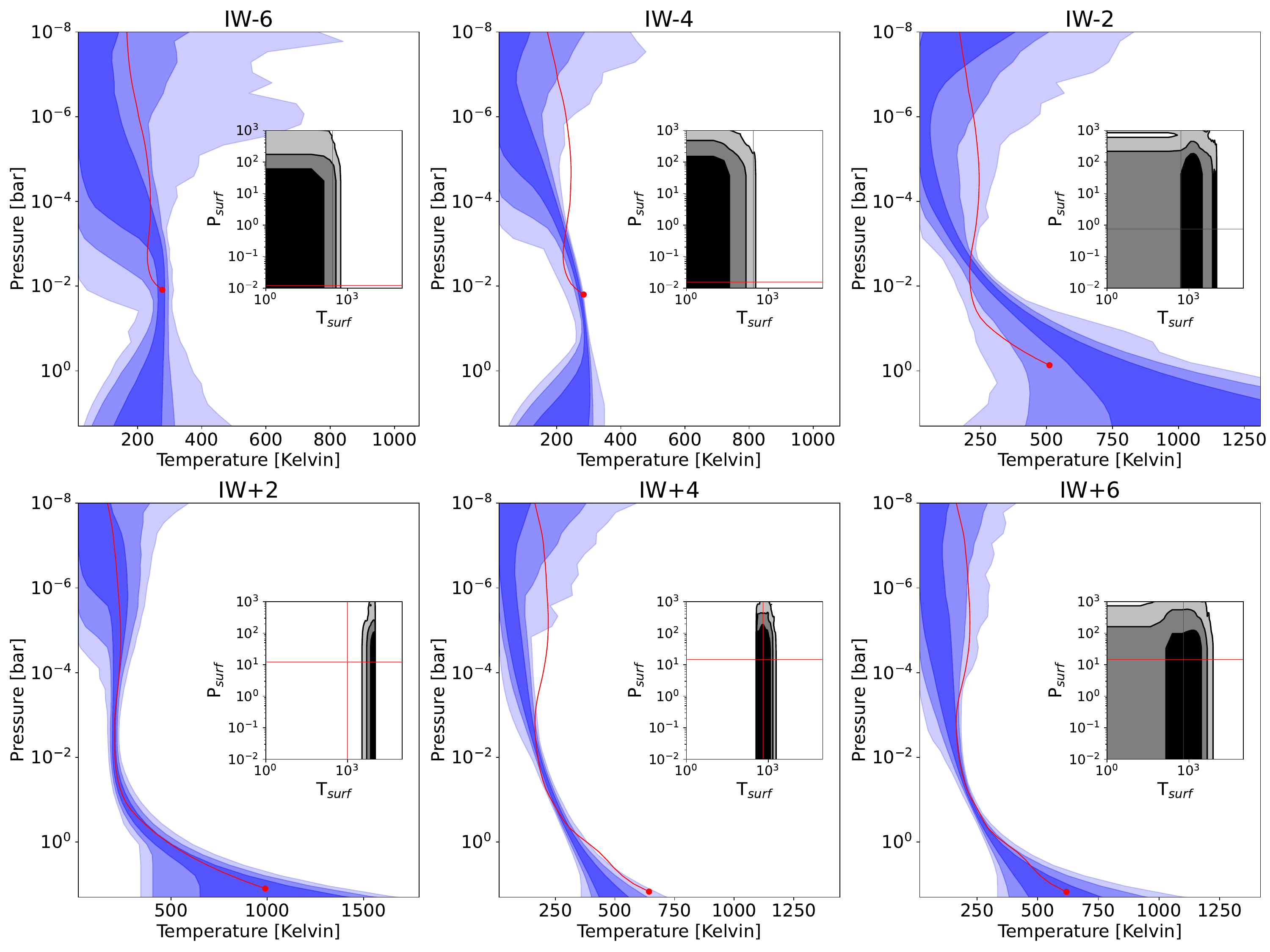}
    \caption{Main plots: Retrieved pressure-temperature ($P$-$T$) profiles (blue regions) with the forward model's PT-profile (red), for each redox state considered (IW-6 to IW+6). The retrieved $P$-$T$ profile's denser region represents the 1$\sigma$ confidence level, intermediate region 2$\sigma$, and thin region the 3$\sigma$ level. The y-axis is cut-off at 20 bar as we consider the opaqueness of the atmosphere. Inset plots: Retrieved joint posterior distributions of surface temperature and pressure (gray) with denser region showing the $\sigma$ confidence levels, as in main plots. Red lines represent the planet's true value.}
    \label{fig:PT-retrievals}
\end{figure*}

\subsection{Emission spectra}

We note from the fluxes alone that the oxidised planets appear to show more pronounced spectral features, with several emission/absorption features observed especially in the 8-14 micron range. While the more reduced cases, such as IW-6 and IW-4, exhibit a comparatively featureless spectrum, IW-2 begins to show more defined features. Our classical retrieval depends strongly on the observed spectrum, thus we expect to be able to retrieve the oxidised atmospheres' composition more accurately as the spectral features provide us with stronger constraints (see Fig. \ref{fig:emission-retrievals}).
It is also worth pointing out that the oxidised planets emit more overall flux and mostly in the center of the observed wavelength region, while the reduced cases' flux increases together with wavelength and peaks at the longer wavelengths of the observational window. \\
Firstly, we compare our forward model emission fluxes with the retrieved emission flux. The results can be seen in Fig.~\ref{fig:emission-retrievals} where the retrieved spectra at 1, 2 and 3 $\sigma$ levels are shown in blue, together with the true planet flux in red. The original flux is recovered within the $\pm$1$\sigma$ level nearly across the whole wavelength range. Except for some slight deviations scattered in the more reduced cases (top row), we mainly observe deviations from the original flux in the longer wavelength regime of the more oxidised planets (especially IW+4 and IW+6). Notably, in the oxidised cases we retrieve the characteristic 15 micron CO\textsubscript{2} feature, especially for IW+2 where the original flux is still within the 1-$\sigma$ bound.

\begin{figure*}
    \centering
    \includegraphics[width=0.85\textwidth]{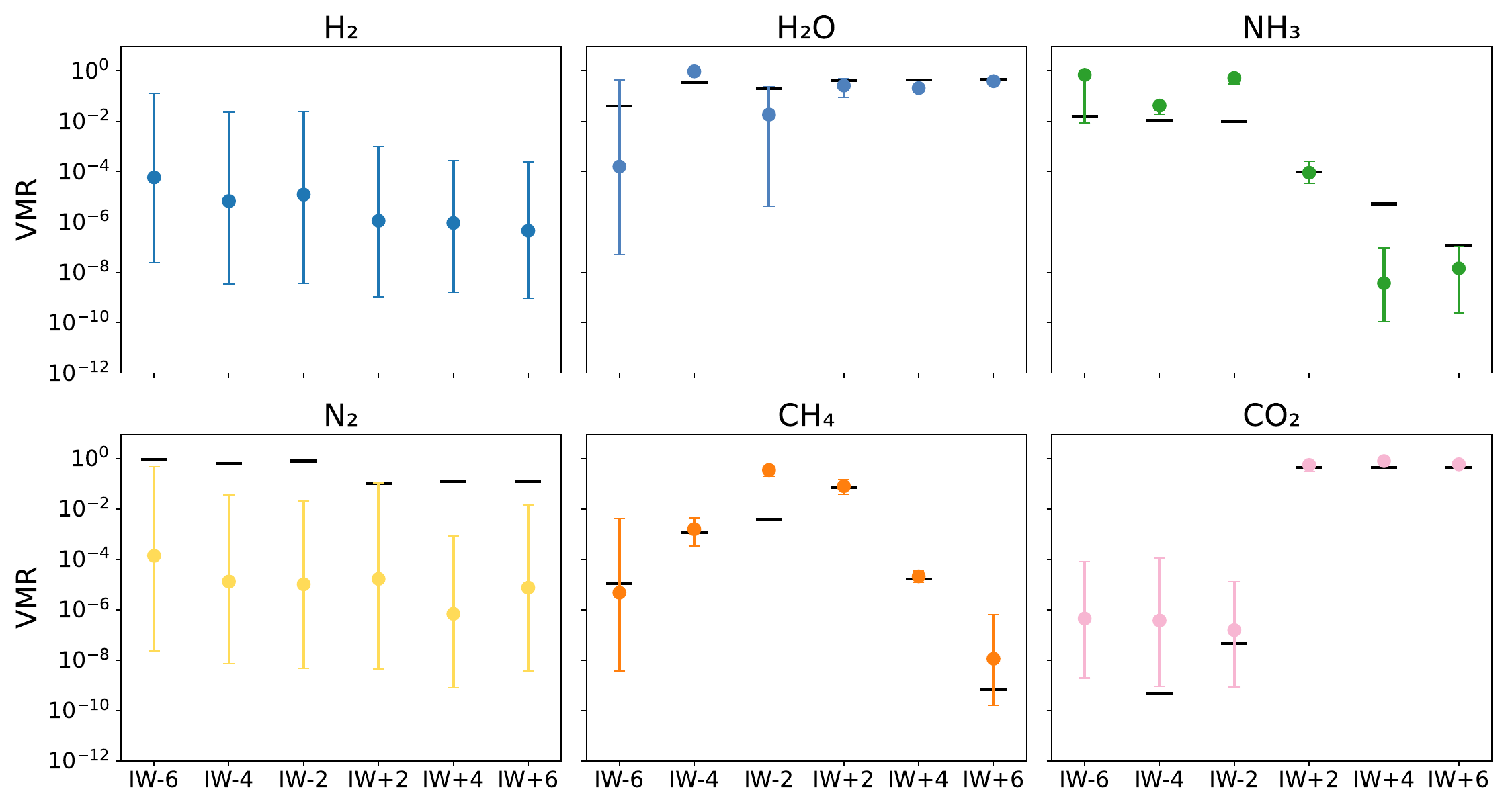}
    \caption{Plot of retrieved volume mixing ratios of each molecule across the six considered redox states. The retrieved median VMR value is shown as a scattered marker, with its corresponding 1$\sigma$ uncertainty. The black horizontal lines indicate the forward model's value. Retrievals register the presence and/or absence of key molecules (i.e., CO\textsubscript{2} and NH\textsubscript{3}) in determining the general oxidized or reduced state of a planet. The trend of CH\textsubscript{4} peaking at intermediate states is captured by the retrieval, making it relevant in distinguishing intermediate states.}
    \label{fig:retrieved_abundances}
\end{figure*}

\begin{figure*}
    \centering
    \includegraphics[width=0.85\textwidth]{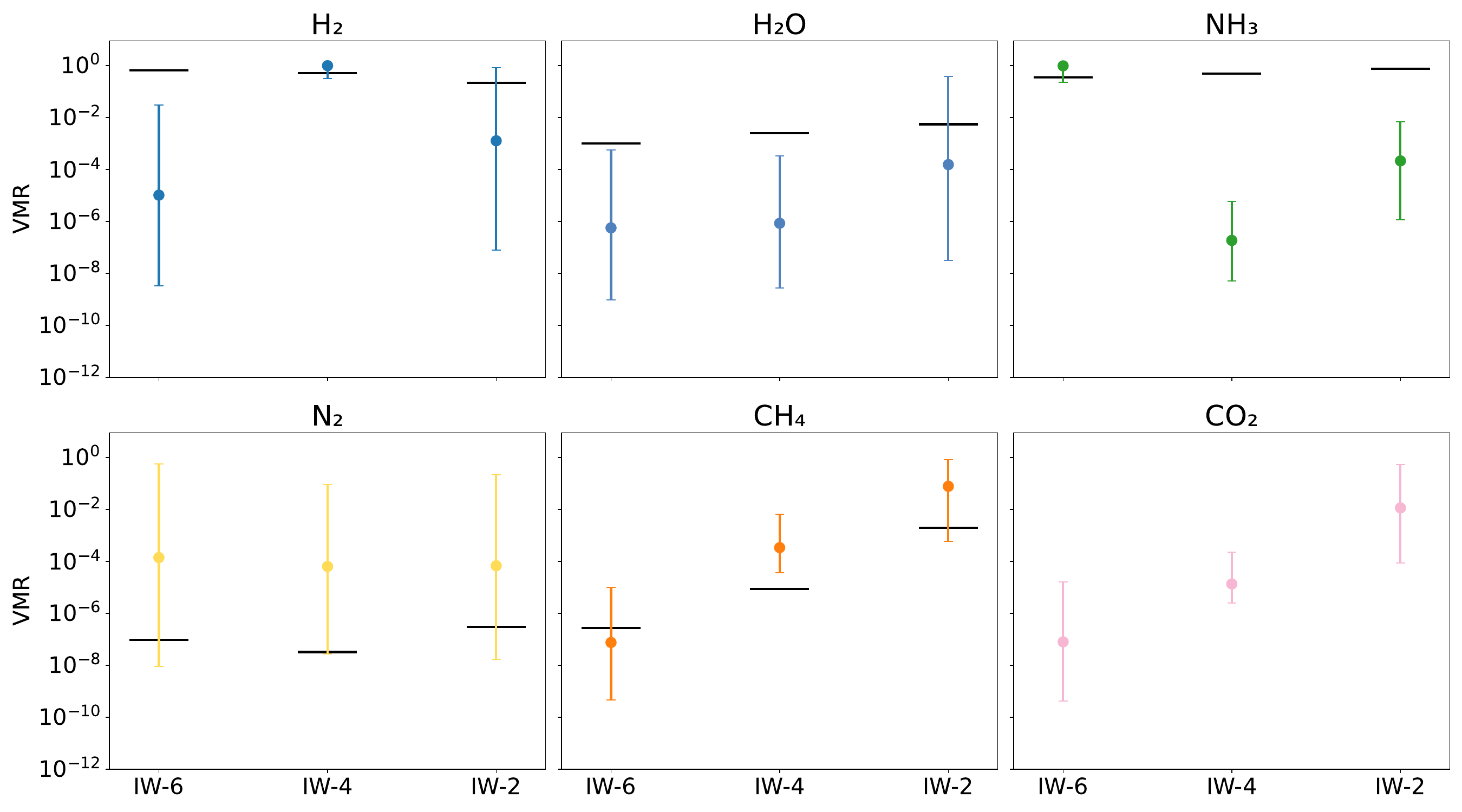}
    \caption{Same as in Fig.~\ref{fig:retrieved_abundances} but for high melt surface flux ($10^{15}$kg yr$^{-1}$) planets with a major H$\textsubscript{2}$ present in the atmospheres of \textit{IW-6, IW-4 and IW-2}. Retrievals of CH\textsubscript{4} and CO\textsubscript{2} mirror previous results, NH\textsubscript{3} is not well retrieved and H\textsubscript{2} appears highly sensitive to factors other than the melt surface flux.}
    \label{fig:H2}
\end{figure*}

\subsection{Pressure-temperature profiles}

On top of the emission flux, with the retrieval we analyse the ability to retrieve the planet's temperature-pressure structure in the atmosphere, down to a pressure limit of 20 bar past which we consider the atmosphere opaque. The results are shown in Fig.~\ref{fig:PT-retrievals}, where the red curve in the main plot is the forward model and the blue regions represent the $\pm 1,2,3 $ $ \sigma$ levels of the retrieved structure. The inset plots show the posterior distributions of the surface temperature and surface pressure (gray), of the respective redox state case, with the straight red lines being the true value of the surface pressure and temperature (i.e., the position of the dot at the end of the red $P$-$T$ curves). Important to note that the resulting temperature from the forward models can differ from the 600K used in the outgassing models (as shown in the vertical red lines). This is due to the nature of the atmospheric model which does not include a radiative-convective calculation of the atmosphere, which is then done in ARCiS for spectra generation. This leads to a difference in surface temperatures of the model exoplanets.

In general, for all six considered redox states, we do not constrain the surface pressure and manage only to retrieve an upper limit, with our confidence levels ranging over several orders of magnitudes. For the reduced environments (top row) especially, the upper level is at least 2 orders of magnitude above the true value in the less reduced case, and multiple orders of magnitude off for the other two (IW-4 and IW-6). For the IW-2 case, the 1$\sigma$ upper limit is two orders of magnitude off and we manage to constrain the surface temperature of the planet with the true value falling at the edge of our 1-$\sigma$ confidence level. For the IW-6 and IW-4 cases, the temperature is unconstrained, but still falls within the $2\sigma$ confidence level for IW-6 and $3\sigma$ for IW-4. In general the retrievals provide us with an upper limit on surface temperature, with only the IW-2 1$\sigma$ level constraining it. Also, the general structure of the $P$-$T$ profile differs for each redox state, with the IW-6 remaining within the $2\sigma$ level, IW-2 within $3\sigma$, and IW-4 being entirely outside of the retrieved profile between $\sim 10^{-4} - 10^{-6} $ bar but within 3$\sigma$ otherwise. \\
For the oxidised planets (bottom row) the situation is quite different. Firstly, we are able to more accurately recreate the $P$-$T$ structure, especially for the IW+2 case where the curve remains consistently in the 1$\sigma$ range. We attribute the performance of this fit to the strong, well-defined spectral features (in particular the 15 micron CO\textsubscript{2} band) shown in the spectrum. The combination of high signal (note the different y-axis scale) and well defined features in the IW+2 case yields the tightest constraints. In the more oxidised cases, IW+4 and IW+6, the structure is more accurately constrained between pressures of $10^{-3}$ and $1$ bar. It then diverges for IW+4 to $3\sigma$ at the surface while IW+6 remains consistently below the $2\sigma$ limit. Secondly, for the surface pressure we always manage to retrieve an upper limit. The $1\sigma$ levels are consistent with the true value for all oxidised planets, i.e. the true values of the planet's surface pressure fall within the $1\sigma$ confidence level. Meanwhile, we manage to constrain the surface temperature for IW+4 and IW+6. In the IW+2 case the surface temperature is overestimated, nonetheless an upper limit on surface pressure is still found.  We attribute the precision of our retrieval within $10^{-1}-10^{-3}$ bar, for oxidised planets, to the fact that these pressures correspond to the region of the atmosphere where an optical depth of $\sim 1$ is reached so these layers contribute most to the observed spectrum. We stress that the $P$-$T$ profile is well recovered only in the photospheric region near optical depth $\sim$ 1 ($\sim$ $10^{-1}-10^{-3}$ bar), which dominates the emergent flux. The physical surface itself lies below the photosphere and is not directly probed by the spectrum; the retrieved surface temperature can therefore be overestimated and is generally weakly constrained (see reduced cases in Fig.~\ref{fig:PT-retrievals}). For IW+2 this leads to a well-fitted photospheric profile alongside an overestimated surface temperature in the joint posterior. This insensitivity of thermal emission to the opaque atmosphere is a phenomenon already known in rocky planets’ retrieval literature \citep{piette2022,Piette2023ApJ}. We therefore plot the retrieved $P$-$T$ profile up to a pressure of 20 bar, as an indicative level where the atmosphere becomes opaque to thermal radiation.

\subsection{Molecular abundances}

Next, we analyse to what extent the molecular abundances in the atmosphere are retrieved. For the corner plots containing the full-fledged retrieval results of our planets, we refer the reader to Appendix \ref{sec:appendixB}. Distinguishing the redox state of a planet with an atmosphere depends to a large extent on what we know of its atmosphere's chemical composition, since the redox state governs the volatiles outgassed in the formation of the planet's secondary atmosphere. The colored scatter points in Fig.~\ref{fig:retrieved_abundances} represent the retrieved Volume Mixing Ratio (VMR), also referred to in the manuscript as molecular abundance, in the atmosphere of the planets. This VMR is obtained by normalizing the retrieved surface partial pressure of each molecule by the total surface pressure (obtained by summing the molecules' partial pressures). The black horizontal lines, in each figure, represent the resulting molecular abundances (i.e. VMR) from the model by \cite{2025Icar..42916450B} (see Fig.~\ref{fig:noack1}) for the respective redox state, while the colored scatter point shows the VMR we retrieve with ARCiS, together with its associated uncertainty, after our observation with LIFE.

In this work, to recreate a real observation scenario to the best of our abilities, we sample the noise discretely along the spectrum and therefore introduce a scatter in the datapoints. The retrieved value is thus expected to vary within the retrieved error bar, this is a statistically expected outcome. While we do not retrieve the accurate surface partial pressure of each molecule in the planet's atmosphere, we retrieve the relative molecular abundances with sufficient accuracy. For example, we successfully retrieve CH\textsubscript{4} (methane) within 1-$\sigma$ in every oxidation state except for IW-2. This reflects its trend of peaking around a neutral IW and decreasing as the environment becomes more oxidising or reducing. Similarly, we retrieve CO\textsubscript{2}'s and H\textsubscript{2}O's abundances in the oxidised cases and within reasonable 1$\sigma$ for IW-2. These are promising results since H\textsubscript{2}O, CO\textsubscript{2} and CH\textsubscript{4} represent key molecular species expected in oxidized planetary atmospheres. Moreover, we obtain satisfying results for NH\textsubscript{3} which is a molecule we expect to find in very reducing conditions, with only IW-2 being off by multiple $\sigma$ levels ($> 5\sigma$).
Nonetheless, we struggle in retrieving N\textsubscript{2}, with the oxidised planets' only retrieving an upper limit to the  N\textsubscript{2} mixing ratio. This we attribute to N\textsubscript{2}'s weak (and/or absent) spectral features in our wavelength ranges, which makes it harder for our observationally constrained retrieval to detect it. Its signatures, if present, are likely too weak compared to the observational noise of LIFE \citep{2022A&A...665A.106A}. Meanwhile, H\textsubscript{2}, while not present in any of these planets, is consistently retrieved below 10$^{-6}$. These first-hand results indicate that we can adequately retrieve key molecules in oxidised planetary environments, with IW+2's retrieval especially in agreement with the model planet.  

On the other hand, our results worsen for the reduced planets. Firstly, the same problems described before persist for H\textsubscript{2} and N\textsubscript{2}, regardless of N\textsubscript{2} being the most abundant specie in the atmosphere. Similarly, H\textsubscript{2}O is within 1 $\sigma$ for IW-6 and IW-2, but the IW-6 is highly unconstrained, i.e. large error bars, (also for CH\textsubscript{4}) while it appears to be overestimated for IW-4. Interestingly, we successfully retrieve CH\textsubscript{4}'s abundance in IW-4, while overestimating it in IW-2 (off by multiple $\sigma$). NH\textsubscript{3} and CO\textsubscript{2} also struggle to be constrained in reduced planets, with CO\textsubscript{2} being almost inexistent in the most reduced case (below lower limit of our retrieved VMR) but nonetheless retrieved at about $10^{-6}$ consistently. This underlying amount is not retrievable as there is essentially no feature in the spectrum caused by CO\textsubscript{2}. This behaviour in our results could be attributed to a combination of the degeneracies in the absorbing regions of CO\textsubscript{2}, CH\textsubscript{4} and H\textsubscript{2}O and the higher noise levels (especially compared to flux magnitude) at lower wavelengths. These results reflect the expected behaviour of H\textsubscript{2} and N\textsubscript{2} molecules with respect to our retrieval framework. N2 and H2 do not have molecular lines that we can directly resolve in the wavelength range we consider. However, they do influence the emission spectrum in three ways: pressure broadening of molecular lines from other molecules, CIA opacity of N2 and H2 themselves, and their influence on the mean molecular weight of the atmosphere as a whole. The retrieval method takes all these effects into account and therefore has some modest constraints on the presence of these molecules. However, these effects are broad and degenerate with other model parameters, which is why it is difficult to pin down their exact abundances or the total surface pressure in case they dominate the atmosphere. In general, while our obtained molecular abundances fluctuate around the true values for reduced planets, we nonetheless obtain results that capture the general underlying trends. In the next sections we will analyse how much resolving power this leads to.

\begin{figure*}
    \centering
    \includegraphics[width=0.85\textwidth]{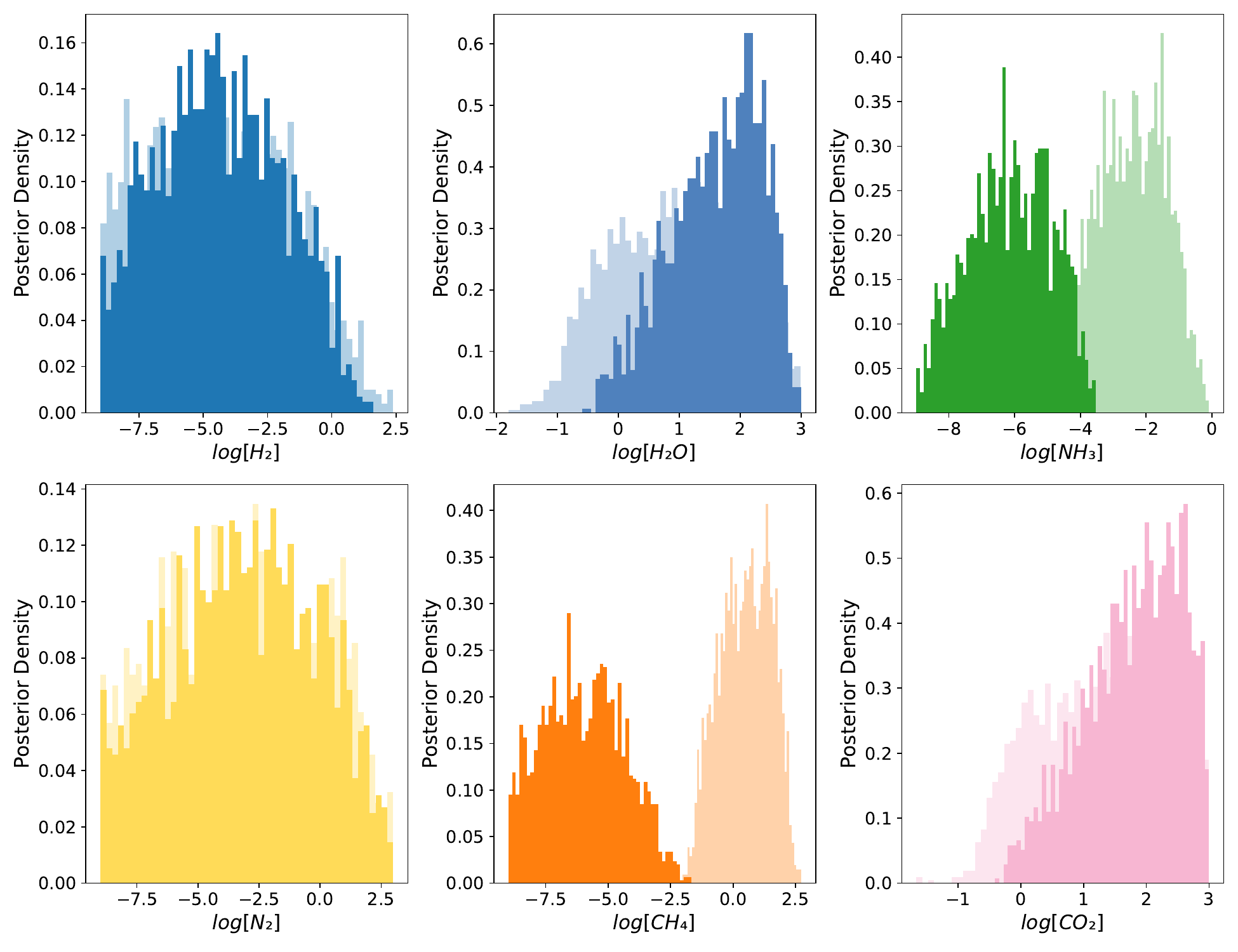}
    \caption{Histograms of the retrieved posterior density distributions of molecular partial pressures of the most oxidised planet IW+6 (opaque) and IW+2 (transparent). Comparing posterior distributions of two oxidised cases highlights the role of key tracers such as methane (CH\textsubscript{4}) in differentiating between oxidised environments without relying on the retrieved abundances of characteristic molecules such as CO\textsubscript{2}/NH\textsubscript{3}.}
    \label{fig:abundances62}
\end{figure*}

\begin{figure*}
    \centering
    \includegraphics[width=0.95\textwidth]{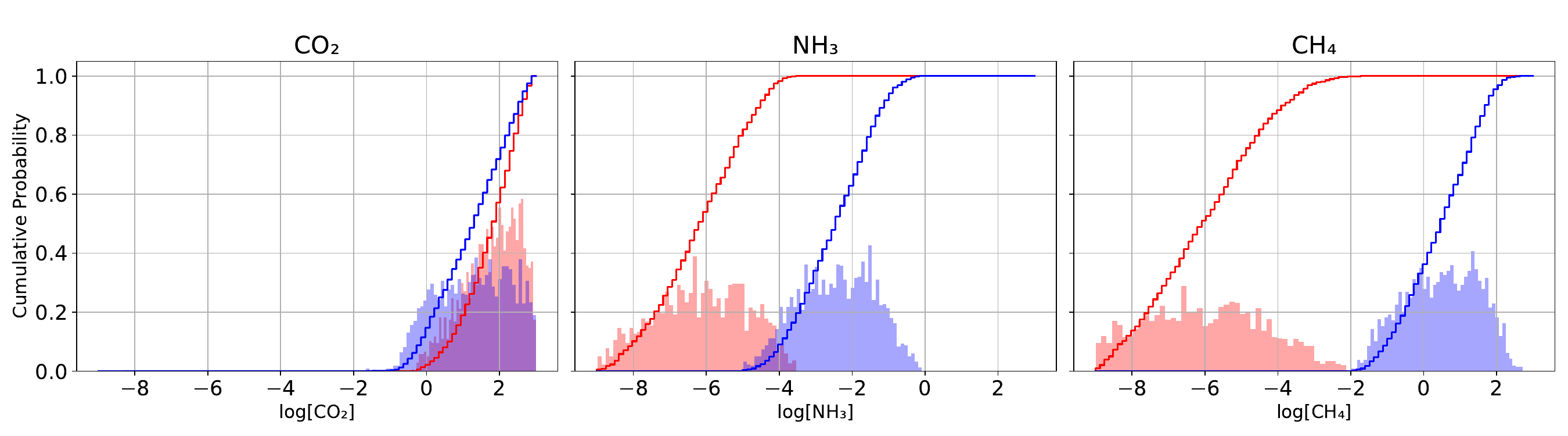}
    \caption{Cumulative probability distributions of CO\textsubscript{2}, NH\textsubscript{3} and CH\textsubscript{4} of the posterior distributions from Fig.~\ref{fig:abundances62} (Red: IW+6; Blue: IW+2). $\Delta$ represents the maximum distance between the cumulative distribution curves, and its value is respectively: 0.284, 0.899, 0.998. The figure shows our statistical method applied to two oxidized environments, highlighting the quantitative measure of redox state differentiation $\Delta$ for key molecules, with emphasis on methane's role.}
    \label{fig:CDF-comparison}
\end{figure*}

\subsubsection*{Retrieving H\textsubscript{2} in high melt surface flux planets}

Our results for H\textsubscript{2}, shown in Fig.~\ref{fig:retrieved_abundances}, prompted us to attempt to constrain the abundances of planets with H\textsubscript{2} in their atmosphere. This is possible using further data from \cite{2025Icar..42916450B} for planets with a higher melt surface flux, defined as the rate at which a planet's interior delivers molten material to the surface \citep{2025Icar..42916450B}. Therefore, a higher flux ($10^{15}$ kg yr$^{-1}$) leads to increased outgassing from the planet's interior, which in turn causes it to release more reduced volatile species, such as H\textsubscript{2}, and surpasses energy-limited hydrogen escape fluxes \citep{2025Icar..42916450B}.
Furthermore, in reduced environments these outgassed volatiles favor the formation of H\textsubscript{2} instead of H\textsubscript{2}O. We show the results of our retrievals on such planets in Fig.~\ref{fig:H2}. We have performed the same process as outlined before for the "standard" planets for these three reduced cases with major amounts of H\textsubscript{2} in their atmosphere. The only difference between these planets and the previous ones, other than the melt surface flux, is that their surface temperature in the model was 300 Kelvin instead of 600, which also resulted in higher surface pressures. The planets with the same melt surface flux, with a surface temperature of 600 Kelvin, had very little (to none) H\textsubscript{2} in their atmospheres. The same is true for the oxidised cases of these same planets.

The results are different than for the reduced cases in Fig.~\ref{fig:retrieved_abundances}. Firstly, while we still retrieve CH\textsubscript{4} relatively well (within 1-2$\sigma$) and detect the almost inexistent CO\textsubscript{2} (as per the reduced cases from Fig.~\ref{fig:retrieved_abundances}). Important to note that the true values of CO\textsubscript{2} are much lower than our minimum retrievable VMR. and thus do not appear in the plot. We fail to constrain H\textsubscript{2}O in all 3 cases, with our distributions showing large error bars, and NH\textsubscript{3}'s retrieval seems erratic (under-retrieved in IW-4 and IW-2, but constrained in IW-6). Despite this, our retrievals of N\textsubscript{2} are slightly improved, with all three cases unconstrained but nonetheless within 1$\sigma$. Finally, H\textsubscript{2} is appropriately retrieved in IW-4, but unconstrained in the other two cases: within 1$\sigma$ in IW-2 and off by almost 6 orders of magnitude in IW-6. While retrieval performance for species like CH\textsubscript{4} and CO\textsubscript{2} mirrors previous results, our retrieval still struggles in constraining H\textsubscript{2}O and NH\textsubscript{3} in these reduced cases. Finally, our retrievals of H\textsubscript{2} vary in performance across redox states, with reasonable agreement in IW-4 and IW-2, but substantial underestimation in IW-6. These results suggest that while H\textsubscript{2} can be retrieved under favorable conditions, its spectral features are still weak and the retrieval accuracy is very sensitive to multiple factors such as surface temperature and mantle redox state.

\subsection{Analysis of posterior distributions}

Having quantified the ability to retrieve the molecular abundances of our target planets, we proceed to assess whether we can distinguish individual redox states from one another. As described in the \hyperlink{methods}{Methods} section, we approach this by computing and comparing the cumulative distribution functions of two redox states, taking the six considered molecules' posterior distributions into account. With a grid of 6 redox states, namely $f$O$_2$= IW[+6,+4,+2,-2,-4,-6], the number of comparisons we can make is large and each case will depend on different combinations of molecules. Every redox state distinction will be case-dependent, nonetheless we expect CO\textsubscript{2} and NH\textsubscript{3} to play a major role in distinguishing oxidised planets from reduced ones. We expect this because they are each produced characteristically in oxidising (CO\textsubscript{2}) and reducing (NH\textsubscript{3}) environments, and because we can detect and resolve their spectral features well enough to be able to retrieve their relative abundances. While we do not retrieve the exact abundance of NH\textsubscript{3}, we nonetheless correctly retrieve its trend of decreasing as the environment becomes oxidised, and vice-versa for CO\textsubscript{2}. Moreover, CH\textsubscript{4} will also play a major role as we find its contribution (in oxidised environments) to peak in moderate ones and decrease as the planet becomes more oxidised, with the same trend in reduced environments (albeit in the opposite direction). We do not expect neither N\textsubscript{2} or H\textsubscript{2} to contribute to our analysis.\\
We include here an example where we analyze the most oxidised case of IW+6 against the slightly oxidised case of IW+2. The collection of molecular posterior distributions of these two cases, plotted against each other, can be found in Fig.~\ref{fig:abundances62}. We find the distributions of H\textsubscript{2}, N\textsubscript{2} and CO\textsubscript{2} to be mostly overlapping, as expected given the results in Fig.~\ref{fig:retrieved_abundances}. H\textsubscript{2}O is slightly different, but not significantly enough, while NH\textsubscript{3} and CH\textsubscript{4} are clearly distinct.\\
At first glance it seems that CH\textsubscript{4} will be the most defining feature distinguishing these two redox states, with NH\textsubscript{3} as second feature. 
We can better observe this in the direct Cumulative Distribution Function (CDF) comparison of CH\textsubscript{4} in Fig.~\ref{fig:CDF-comparison} with a $\Delta$ value of 0.998 in the range of [0,1]. We also include the CDF comparison of CO\textsubscript{2} (Fig.~\ref{fig:CDF-comparison}) with $\Delta=0.284$, which is not useful in this case as both planets have similar posterior distributions of CO\textsubscript{2}. This indicates that CO\textsubscript{2} abundance alone is not decisive for distinguishing among two oxidised states. Nonetheless, it can be be useful for comparing oxidised atmospheres to reduced ones. When performing this analysis we must keep in mind that we found more CH\textsubscript{4} than in the model, meaning that its real $\Delta$ would be a bit lower than 0.998. Nonetheless, we also must consider that NH\textsubscript{3} is under-estimated, meaning that its true $\Delta$ should be higher. This indicates that we should not rely on a single molecule's distribution, but on combinations of such molecules which can become case-dependent indicators of different underlying redox states.\\ \begin{figure*}
    \centering
    \includegraphics[width=0.95\textwidth]{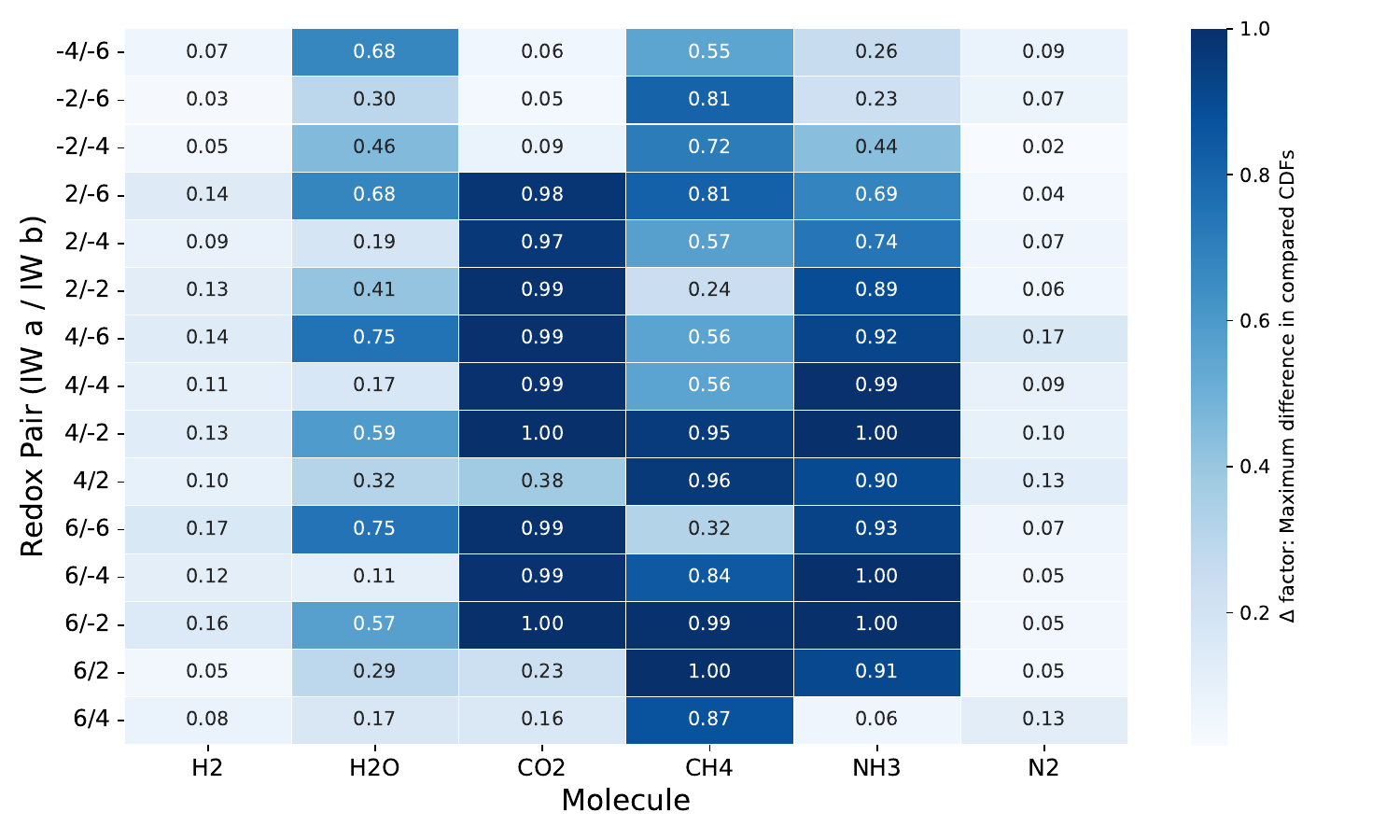}
    \caption{Grid of the final $\Delta$ values comparing each molecule between all available redox states. Y-axis represents the redox states being compared and the x-axis the molecule considered. A high $\Delta$ value indicates more distinct posterior distributions (as per Fig.~\ref{fig:CDF-comparison}) and a higher confidence in attributing different underlying redox states to a molecule. While H\textsubscript{2} and N\textsubscript{2} do not contribute to our results, combinations of key tracers such as CO\textsubscript{2}, NH\textsubscript{3} and CH\textsubscript{4} cover most of the mantle redox state range.}
    \label{fig:Deltagrid}
\end{figure*}
Finally, we compute a grid of $\Delta$ values by comparing all CDFs of the six molecular parameters, across all redox states. The final grid is found in Fig.~\ref{fig:Deltagrid}, the results give an overview of the role of each molecule in the redox state distinction between model planets used in this work. Firstly, retrievals of N\textsubscript{2} and H\textsubscript{2} should not be considered in our analysis, especially if one compares the retrieved values to the ground truth. Secondly, H\textsubscript{2}O does not play as much of a role as one might expect at first. It reaches slightly relevant, but not conclusive, $\Delta$ values when comparing the most reduced planet (IW-6) to oxidised states, and is nonetheless overshadowed by either (or both) CO\textsubscript{2} and NH\textsubscript{3} in those cases. Further, we find that combinations of CO\textsubscript{2}, CH\textsubscript{4} and NH\textsubscript{3} can provide strong indications of different underlying redox states between two planets across all considered redox states. Except for comparisons among reduced planets where no molecule (or combinations thereof), with the exception of CH\textsubscript{4} between IW-2 and IW-6 ($\Delta=0.81$), can yield a significant result. We find that CO\textsubscript{2} with NH\textsubscript{3} (CH\textsubscript{4} as well in some but not all cases) are most useful in distinguishing oxidised from reduced planets, as they appear to be the molecules most indicative of oxidising conditions in secondary atmospheres. Meanwhile, CH\textsubscript{4} appears to be most important in distinguishing oxidised states from each other, possibly in combination with NH\textsubscript{3}, for IW+6 / IW+2.

\section{Discussion}

\subsection{Constraining the abiotic baseline with LIFE}

Exoplanetary science covers the full range of evolutionary stages of rocky exoplanets, revealing a variety of chemical compositions and physically viable system architectures \citep{wordsworth2022}. While the Solar System’s record is patchy, due to billions of years of dynamical and geological evolution that have overwritten much of the original evidence, observations of exoplanets at different evolutionary stages can help fill in these gaps \citep{lichtenberg2025}. Characterisation, through direct imaging, of regimes analogous to the prebiotic Earth can offer crucial insight into the formation and evolution of abiotic and prebiotic atmospheres, the chemical origins of life, and the distinction between habitable and uninhabitable environments. Establishing a solid foundation of abiotic processes is therefore essential for future searches of extrasolar life through biosignatures in the atmospheres of rocky planets \citep{2022NatAs...6..189K,Seager2025arXiv}.

In this work we present preliminary results that provide a framework for constraining the abiotic baseline of secondary atmospheres. Using LIFE observations and ARCiS retrievals, we are able to recover key molecular abundance trends for Earth-sized planets at 10 pc. To study the observability and atmospheric detection of target molecules we used the LIFE's mission simulator, LIFEsim. The simulator models astrophysical noise including contributions from stellar leakage, exozodiacal dust, and photon noise. We used LIFEsim as the simulator for the astrophysical noise that originates from observing these planets at a distance of 10 parsecs, orbiting a Sun-like star at 1 AU. By integrating until we reach a target S/N of 10 we obtain a planetary spectrum that reflects realistic observational uncertainties. Our approach introduces wavelength-dependent noise via white Gaussian noise (WGN) realizations, emulating the stochastic nature of photon noise in space-based observations.\\
\indent Firstly, we find that the LIFE instrument is capable of distinguishing the redox state of our target planets in most cases. After observations with LIFE and retrievals with ARCiS we can confirm the presence (or absence) of several characteristic molecules in oxidised environments (e.g. CO\textsubscript{2}, CH\textsubscript{4} and H\textsubscript{2}O), with many cases being retrieved within 1$\sigma$ of their true abundance (especially in the moderately oxidised IW+2). We are also able, in most other cases, to detect chemical trends between different redox states by adequately detecting and retrieving relative abundances of the major molecules in their atmospheres (e.g. NH\textsubscript{3}, CO\textsubscript{2} and CH\textsubscript{4}). Our analysis, through a statistical test based on the cumulative distribution of the posterior distributions of molecular partial pressures in the target planet's atmosphere, demonstrates that LIFE's current noise levels, spectral resolution, and wavelength coverage, are sufficient to distinguish redox states under current assumptions. Especially when one relies on a combination of characteristic molecules, such as CH\textsubscript{4} and CO\textsubscript{2}, rather than a single tracer. This is result is corroborated by \cite{Drant2025}, who demonstrated that the CO\textsubscript{2}/CH\textsubscript{4} is a tracer for redox state in warm exoplanet atmospheres. While in our simulations methane plays a major role in both reducing and oxidising conditions, in \cite{Drant2025} they expect the abundances of methane in oxidising conditions to be too small due to photochemistry effects. Furthermore, our work hinges on a future population-level survey of the LIFE instrument to build on JWST observations, which are also expected to provide insight on these mechanisms in rocky exoplanet atmospheres \citep{Drant2025}. Moreover, our framework constrains environments rich in CH\textsubscript{4} and NH\textsubscript{3}. This has implications for feedstock molecules such as HCN (hydrogen cyanide) which is essential in the production of life's building blocks, and was shown to depend upon the presence of carbon and nitrogen-bearing species (i.e. CH\textsubscript{4} and NH\textsubscript{3}) in Earth-like rocky planets \citep{2019Icar..329..124R}.\\

Our results indicate that with LIFE, by retrieving atmospheric compositions across a range of redox states, it would be possible to test whether observed spectra are consistent with oxidized, Earth-like atmospheres or more reducing regimes. Distinguishing these types of planetary environments is critical to understand the diversity in the exoplanet census and to establish whether Earth-like prebiotic chemistry pathways would be possible on terrestrial exoplanets. Further, we can confidently distinguish between observations of oxidising or reducing environments in the observed exoplanets. Future observations with LIFE may thus shed light on the mechanisms driving the origin and evolution of secondary atmospheres. The observable redox state, through emission spectroscopy, acts as interface between the interactions of a planet’s interior and its atmosphere: the mantle’s internal evolution critically influences the volatile inventory available to the atmosphere, in turn these atmospheric gases regulate heat loss to space, and thus impact the planet's evolution on billion-year timescales \citep{Lichtenberg2025Science}. Therefore, future robust retrievals of key molecules with LIFE may enable to discern abiotic chemical environments from prebiotic chemistry, via constraints on CO\textsubscript{2}/CH\textsubscript{4} abundances and/or on species favourable to prebiotic feeding blocks (CH\textsubscript{4}/NH\textsubscript{3}) such as HCN for example, and possibly provide evidence towards biosignature detections. Our results therefore show that characterisations of exoplanets' redox states with LIFE can provide a solid framework for direct imaging characterisation of exoplanet atmospheres to test and constrain theoretical models of exoplanetary evolution and thus inform viable prebiotic environments on these worlds.

\subsection{Limitations and future work}

In this work we made a number of assumptions that constrain the generalizability of our results. Firstly, we assumed the atmosphere of our planets to be dominated by a fixed set of six molecules: H\textsubscript{2}, H\textsubscript{2}O, CO\textsubscript{2}, CH\textsubscript{4}, NH\textsubscript{3}, and N\textsubscript{2}. While we do expect these species to be prevalent in planets' secondary atmospheres that form through processes such as (volcanic) outgassing, real atmospheres will have other molecules that also shape their emission flux. Further, they may also be shaped by other processes that are not taken into consideration within this work (e.g. clouds, photochemistry). For example, we expect clouds to cause biases in the thermal structure of the atmosphere, although cloud-free retrievals still yield accurate results on the atmospheric composition. Finally, our simulations do not take into account potential effects of photochemistry. Therefore, the abundances of NH$_3$ and CH$_4$ may be affected in thermal regime below $\approx$700 K \citep{liggins2023,Drant2025}. In future work we aim to incorporate these effects, in addition to time-variable outgassing and self-consistent atmospheric evolution. \\

Secondly, to limit the complexity of our work we assumed an ideal relationship between the mixing ratios and partial pressures of the molecules in our model. This allowed us to assume that each parameter could be independently retrieved without stringent chemical or physical constraints. While giving us maximum flexibility in the retrieval space, this approach may also introduce degeneracies in our findings and could result in physically inconsistent solutions, especially for species that we cannot constrain from flux observations with LIFE. Similarly, we retrieve a free temperature profile which does not impose parametric or equilibrium constraints (enhancing realism) but increases the complexity and required computing power of the retrieval.\\
A further limitation is that the atmospheric composition is taken directly from the degassing model of \citet{2025Icar..42916450B} and is vertically constant along pressure, when generating the emission spectra. We do not re-equilibrate the gas speciation to the radiative-convective $P$-$T$ profile subsequently computed by ARCiS. We do not couple the radiative effect back onto the surface temperature and this can lead to biases in our results. This is especially the case in reducing conditions, where methane can build up, in which strong climate feedbacks are expected \citep{Drant2025}. A self-consistent coupling of degassing, atmospheric chemistry, and a radiative–convective climate calculation is left to future work. \\
\indent Moreover, our statistical framework is relatively simplistic. The statistical test used in this work relies on the $\Delta$ metric to quantify redox state distinction. While robust and useful for first-hand intuitive results, it does not take into account covariance and correlations between molecules that could provide a more comprehensive overview. For future analysis, we encourage the development of a more complex metric, or the possibility of a machine learning approach, to better capture the strength and limitations of our, and future, redox state distinctions in rocky exoplanets. \\
\indent Finally, this study was constrained by practical matters such as keeping a manageable parameter space and computing power for the retrievals. Thus we limited our measurements to a fixed stellar type, planet orbital radius, planetary characteristics (Earth-like mass and radius), and distance of observations. We also adopted a fixed instrumental configuration, with observations fixed at a total S/N of 10 and a single dust level, with instrument architecture based on the latest LIFE design and capabilities. We also treated a limited range of planetary surface temperatures and redox states, performing a total of 9 full retrievals. While this study is sufficient for a proof-of-concept, we cannot yet generalize our results to a wider population of exoplanets. \\

Despite these limitations, our results present a proof-of-concept framework that highlights LIFE's capabilities, together with a planetary modeling framework such as ARCiS, in distinguishing the redox state of exoplanets' secondary atmospheres. The use of the same framework (ARCiS) in both forward modeling and retrieval ensures consistency between the spectra generation in forward models and in the retrievals. This avoids problematic systematic discrepancies highlighted in previous works with LIFE and retrievals \citep{2022A&A...665A.106A}. Further, retrieving all atmospheric and planetary parameters as free variables also allows for our retrievals to be unbiased towards the planet's redox state and chemical composition. Finally, we modeled planets from the latest state-of-the-art planetary atmosphere models \citep{2025Icar..42916450B}, allowing us to validate results against a ground truth that provides us a direct and intuitive validation of our retrieval's performance.

Future work should aim to relax some of the assumptions listed above. This refers to a possible expansion of the chemical parameter space, an exploration of other host stars, observational settings, and different target planet's characteristics such as a broader range of surface temperatures in the atmospheric models. A more complex statistical analysis, perhaps through machine learning models, can also provide a more robust analysis of redox state distinction based on our observations with LIFE. Such developments will develop and expand our understanding of LIFE's performance in redox state distinction of exoplanets.

\section{Conclusions}
\label{sec: Conclusion}

In this study we investigated the ability of the LIFE mission concept to distinguish the oxidation states of Earth-like exoplanets via mid-infrared emission spectroscopy and the ARCiS Bayesian retrieval framework. Our main findings can be summarized as follows:

\begin{itemize}
    \item LIFE is capable of retrieving molecular abundances in oxidised atmospheres (e.g. IW+2, IW+4, IW+6), better than in reduced ones, with more accurate surface pressures, temperature profiles, and molecular abundances recovered within 1-2 orders of magnitude. Planets with reduced atmospheres yield relatively flatter spectra with weaker spectral features, which negatively affects observations and increases the uncertainty of our retrievals.

    \item CO\textsubscript{2}, CH\textsubscript{4}, and NH\textsubscript{3} are the key molecules in redox state distinction; we manage to capture strong trends in their distribution across the redox spectrum. CO\textsubscript{2} dominates in oxidised cases, while NH\textsubscript{3} dominates in reduced environments. CH\textsubscript{4} serves as a reliable intermediate tracer. These trends are generally captured by the retrievals and confirmed statistically. These constraints may also serve as tracers for individual redox state retrieval and HCN favourable prebiotic environments.

    \item H\textsubscript{2} detection is challenging and context-dependent. For cooler reduced planets with high melt surface flux, H\textsubscript{2} becomes a major component of the atmosphere. However, retrieval accuracy seems to depend greatly on the surface temperature and redox state of the planet. H\textsubscript{2} is only appropriately retrieved (within 2 orders of magnitude) in moderately reduced cases (e.g. IW-2, IW-4).

    \item Using simulated observations with LIFEsim and retrievals with ARCiS, we show that LIFE’s wavelength coverage (6--16~$\mu$m), spectral resolution (R=100), and noise level are sufficient to retrieve molecular abundances mirroring reality and identify redox-sensitive chemical trends. These are better confirmed particularly when multiple species are considered in combination (such as CO\textsubscript{2} and NH\textsubscript{3} between oxidised and reduced cases and CH\textsubscript{4} among each case).

    \item Our assumptions in modeling and retrieval impact the generalization of our results. The results obtained in this work are limited by restricted model parameters (host star, planet mass, radius, orbit, and distance), a fixed instrumental set-up, and limited chemical environment. While suitable for a controlled proof-of-concept, future work will need to expand this framework to properly address statistical trends across planetary atmospheres and planetary systems.
    
\end{itemize}

In summary, the LIFE mission concept is capable of distinguishing oxidation states of individual exoplanets through observations and characterization of their secondary atmospheres, which are a result of planetary processes such as volatile outgassing and atmospheric escape. In this work we establish a proof-of-concept framework for redox state distinction through the LIFE mission. We recover the redox-sensitive atmospheric information of such planets and distinguish their oxidising and (or) reducing states. Further studies in this field would greatly contribute to our collective understanding of secondary atmosphere formation and evolution on terrestrial protoplanets. These results motivate further retrieval development with LIFE as the observational tool, including broader parameter space, alternative planetary scenarios, and more advanced statistical classification methods.

\section*{Acknowledgments} 
T.L. and L.C. were supported by the Branco Weiss Foundation, the Netherlands eScience Center (PROTEUS project, NLESC.\-OEC.\-2023.\-017), the Alfred P. Sloan Foundation (AEThER project, G-2025-25284), NASA’s Nexus for Exoplanet System Science research coordination network (Alien Earths project, 80NSSC21K0593), and the NWO NWA-ORC PRELIFE Consortium (NWA.1630.23.013). Part of this work has been carried out within the framework of the National Centre of Competence in Research PlanetS supported by the Swiss National Science Foundation under grants 51NF40\_182901 and 51NF40\_205606. SPQ acknowledges the financial support of the SNSF. EA's work was supported by appointments to the NASA Postdoctoral Program at the NASA Goddard Space Flight Center, administered by Oak Ridge Associated Universities under contract with NASA (ORAU-80HQTR21CA005).

\section*{Data Availability}

All code used to obtain these results, along with the data directories, are openly available at \url{https://github.com/lorlo9999/LIFE_RedoxStateCharacterization}.

\hypersetup{
    allcolors=black
}
\bibliographystyle{mnras}
\bibliography{Ref}

\appendix
\label{sec: appendixA}

\section{Radiative Transfer Parameters}

\onecolumn

\renewcommand{\arraystretch}{0.7}
\begin{table*}
\centering
\caption{Table of ARCiS parameters used for Radiative Transfer models}
\begin{tabular}{l|l|l}

\textbf{Parameter}         & \textbf{Value}                                                                                        & \textbf{Description}                                                                     \\ \hline
Tstar             & 5777 {[}Kelvin{]}                                                                            & Temperature of host star.                                                       \\ \hline
Rstar             & 1 [R$_{\odot}$]                                                          & Radius of host star.                                                            \\ \hline
Dplanet           & 1 {[}AU{]}                                                                                   & Orbital distance of planet to its host star.                                    \\ \hline
distance          & 10 {[}pc{]}                                                                                  & Distance star-planet system from the Sun.                                       \\ \hline
Rp                & 1 [R$_{\oplus}$]                                                          & Radius of the planet.                                                           \\ \hline
Mp                & 1 [M$_{\oplus}$]                                                          & Mass of the planet.                                                             \\ \hline
lmin              & 0.1 [$\mu$m]                                                               & Minimum wavelength of spectrum.                                                 \\ \hline
lmax              & 20 [$\mu$m]                                                               & Maximum wavelength of spectrum.                                                 \\ \hline
specres           & 200                                                                                          & Spectral resolution of spectrum.                                                \\ \hline
pmin              & 10$^{-8}$ [bar]                                                                  & Minimum pressure considered in the atmosphere.                                  \\ \hline
pmax              & see Fig. ~\ref{fig:noack2}                                                                                    & Maximum pressure considered in the atmosphere.                                  \\ \hline
Pp                & same as pmax                                                                                 & Atmospheric pressure corresponding to radius Rp.                                \\ \hline
scattering        & True                                                                                         & Logical determining if scattering of the thermal radiation is included.         \\ \hline
scattstar         & True                                                                                         & Logical determining if scattering from the star is included.                    \\ \hline
computeT          & True                                                                                         & Logical determining if the temperature structure is computed self-consistently. \\ \hline
CIA               & \begin{tabular}[c]{@{}l@{}}CO\textsubscript{2}-CO\textsubscript{2}\\ N\textsubscript{2}-N\textsubscript{2}\\ N\textsubscript{2}-H\textsubscript{2}O\\ CO\textsubscript{2}-CH\textsubscript{4}\\ CO\textsubscript{2}-H\textsubscript{2}\\ H\textsubscript{2}-CH\textsubscript{4}\end{tabular} & Collisionally Induced Absorption (CIA) lines considered in radiative transfer   \\ \hline
\textbf{Partial Pressures} &                                                                                              &                                                                                 \\ \hline
H\textsubscript{2}                & see Fig. ~\ref{fig:noack1}                                                                                    & Partial pressure of H\textsubscript{2} molecule in secondary atmosphere of the planet.          \\ \hline
H\textsubscript{2}O               & see Fig. ~\ref{fig:noack1}                                                                                & Partial pressure of H\textsubscript{2}O molecule in secondary atmosphere of the planet.         \\ \hline
NH\textsubscript{3}               & see Fig. ~\ref{fig:noack1}                                                                                     & Partial pressure of NH\textsubscript{3} molecule in secondary atmosphere of the planet.         \\ \hline
N\textsubscript{2}                & see Fig. ~\ref{fig:noack1}                                                                                     & Partial pressure of N\textsubscript{2} molecule in secondary atmosphere of the planet.          \\ \hline
CH\textsubscript{4}               & see Fig. ~\ref{fig:noack1}                                                                                     & Partial pressure of CH\textsubscript{4} molecule in secondary atmosphere of the planet.         \\ \hline
CO\textsubscript{2}               & see Fig. ~\ref{fig:noack1}                                                                                     & Partial pressure of CO\textsubscript{2} molecule in secondary atmosphere of the planet.         \\ \hline
\end{tabular}
\label{tab:radiativetransfer}
\end{table*}

\hypersetup{
    allcolors=blue
}
\begin{figure}
\centering
\begin{minipage}[t]{0.48\textwidth}
\centering
\includegraphics[width=\linewidth]{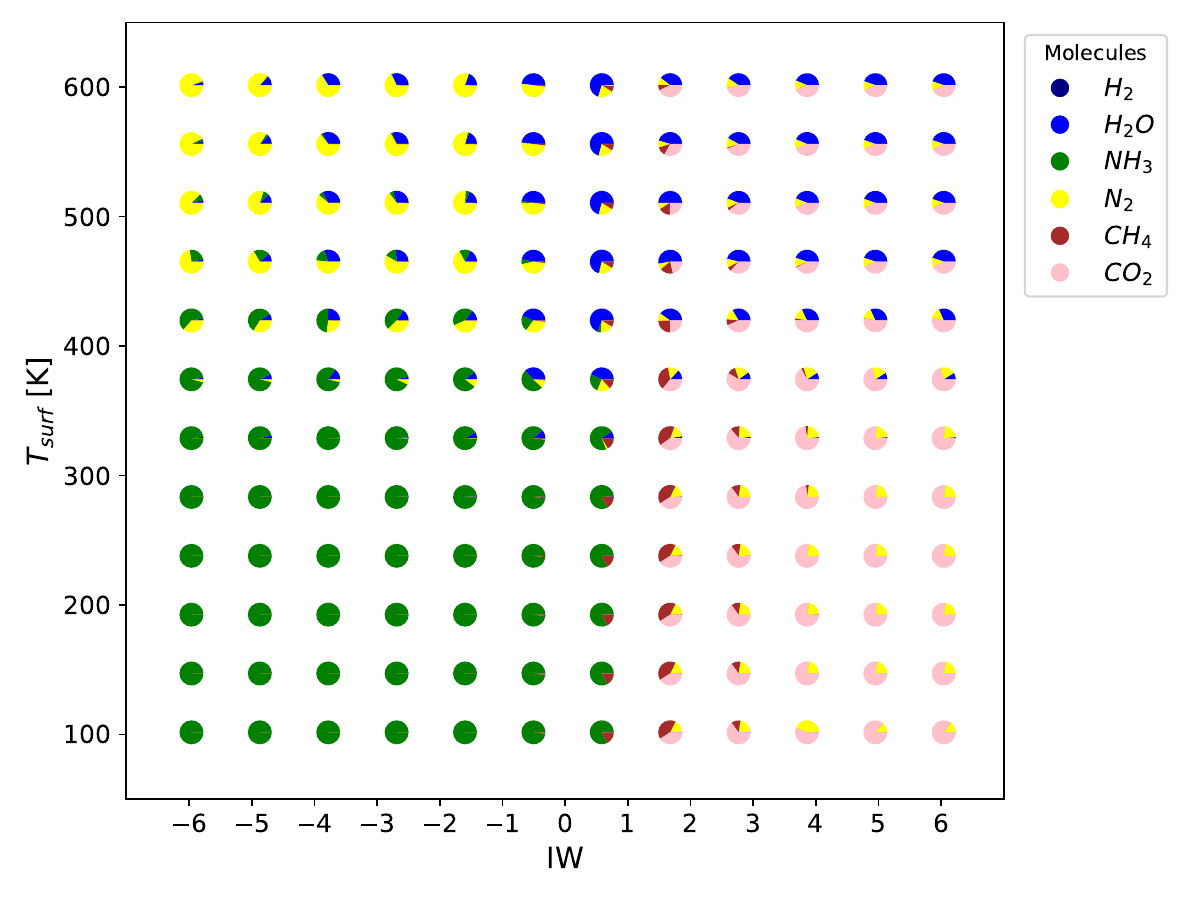}
\caption{Replot from model data of Figure $2$ (a) from \citet{2025Icar..42916450B} showing the resulting molecular volume mixing ratios in the secondary atmosphere of planets at different surface temperatures and redox states. Top row at $600$K includes the planets used for our forward models across the redox spectrum. All atmospheres in \citet{2025Icar..42916450B} are obtained from an initial volatile content equal to post-solidification Earth: of 450 ppm H2O, 50 ppm CO2 and 10 ppm N2.}
\label{fig:noack1}
\end{minipage}
\hfill
\begin{minipage}[t]{0.48\textwidth}
\centering
\includegraphics[width=\linewidth]{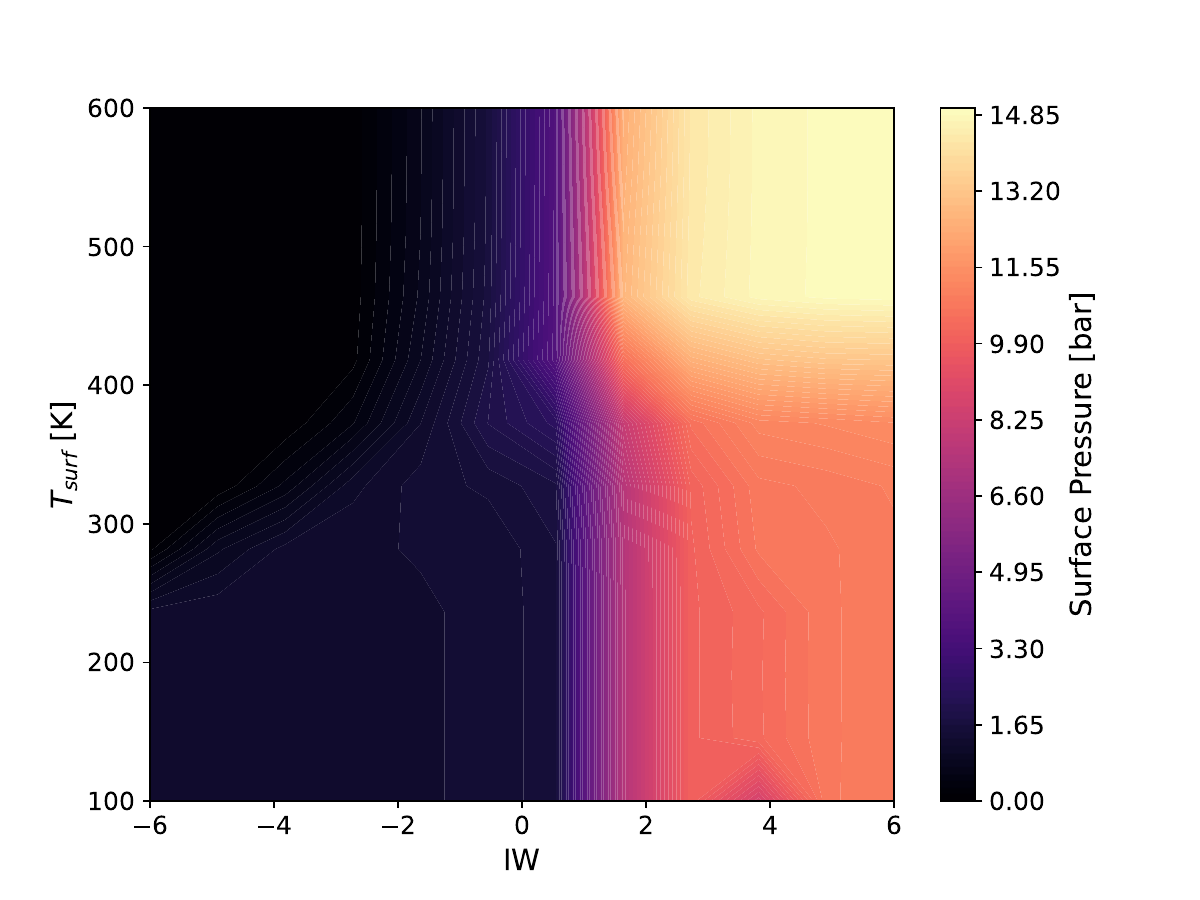}
\caption{Replot from model data of Figure $2$ (b) from \citet{2025Icar..42916450B} showing the resulting overall (surface) pressure of the same planets in Fig.~\ref{fig:noack1}, used in our forward models.}
\label{fig:noack2}
\end{minipage}
\end{figure}

\hypersetup{
    allcolors=black
}

\clearpage
\section{Corner Plots}
\label{sec:appendixB}

\begin{figure*}
    \centering
    \includegraphics[width=1\linewidth]{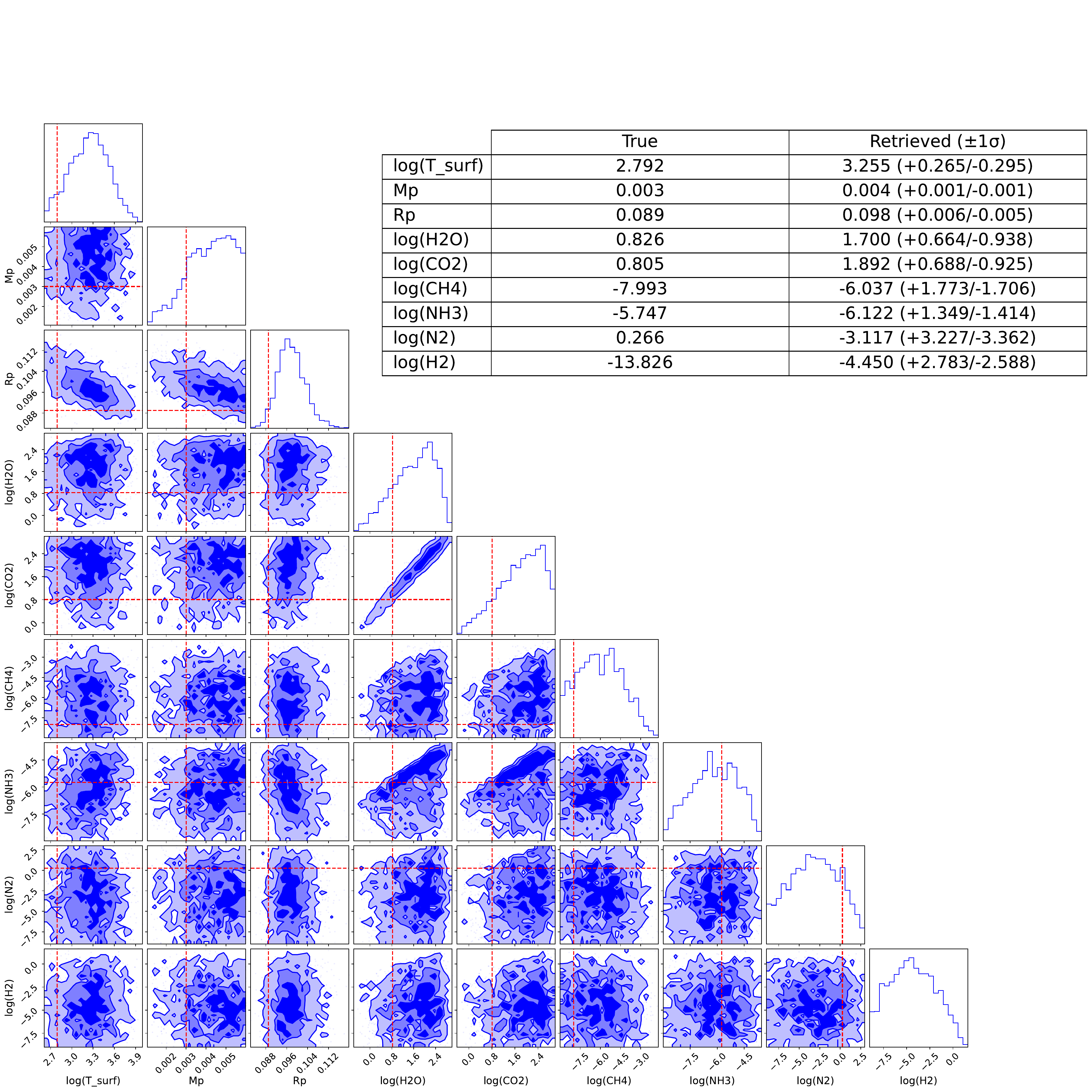}
    \caption{Corner plot of the posterior distributions from retrievals on the IW+6 redox state case. The red lines indicate the true values of the planet. The table in top-right corner includes the true values for direct comparison.}
    \label{fig:cornerIW+6}
\end{figure*}

\begin{figure*}
    \centering
    \includegraphics[width=1\linewidth]{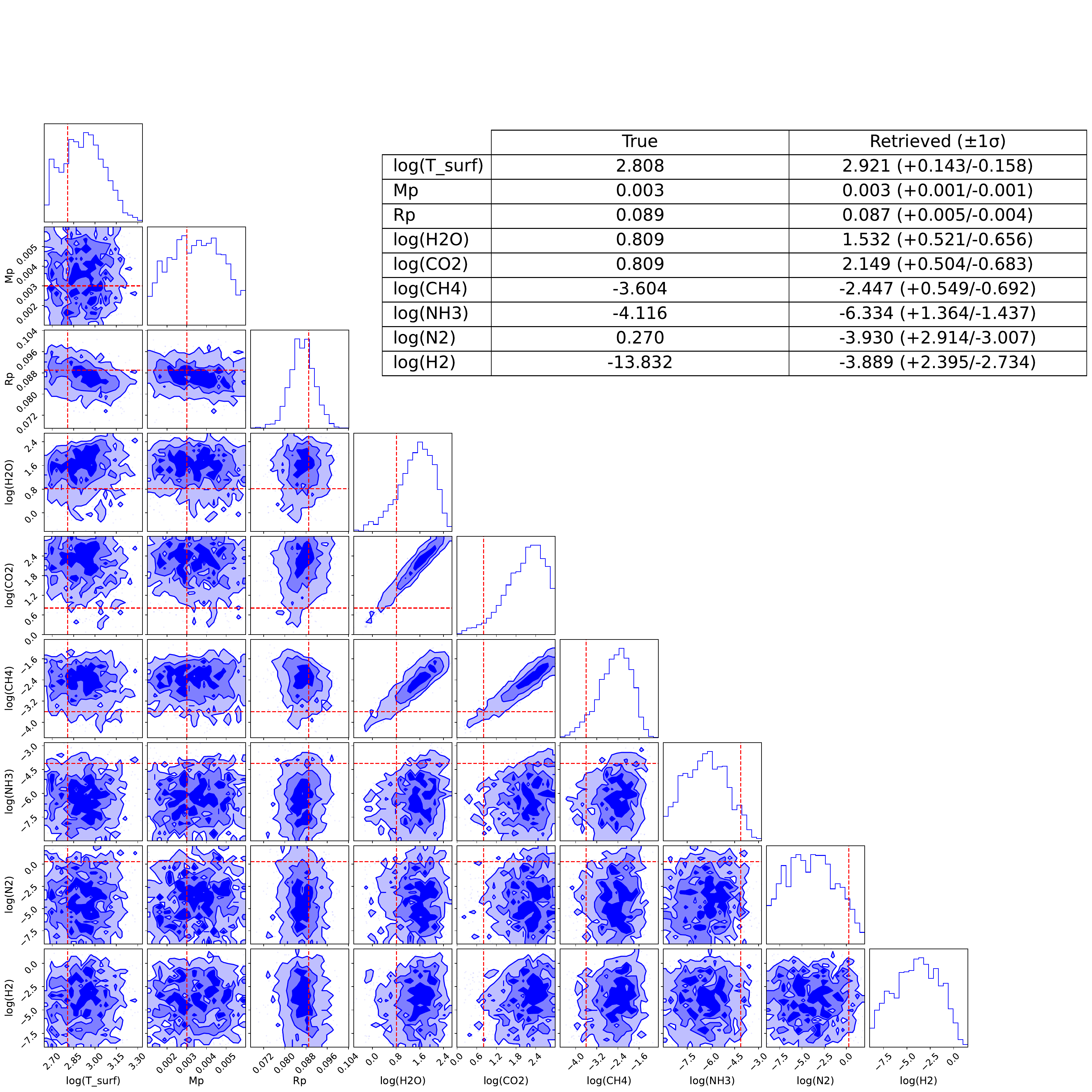}
    \caption{Same as in Fig.~\ref{fig:cornerIW+6} but for IW+4}
    \label{fig:cornerIW+4}
\end{figure*}

\begin{figure*}
    \centering
    \includegraphics[width=1\linewidth]{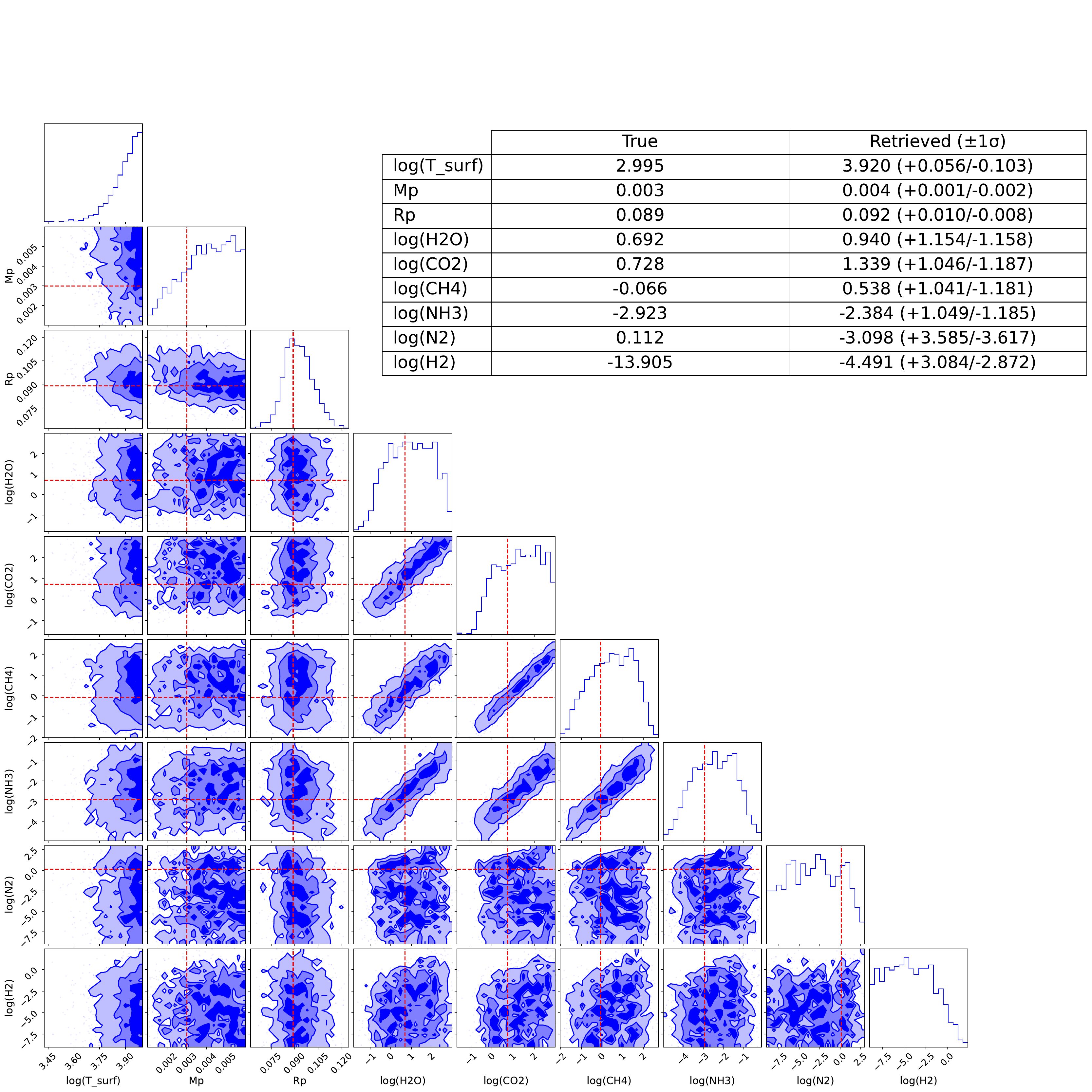}
    \caption{Same as in Fig.~\ref{fig:cornerIW+6} but for IW+2}
    \label{fig:cornerIW+2}
\end{figure*}

\begin{figure*}
    \centering
    \includegraphics[width=1\linewidth]{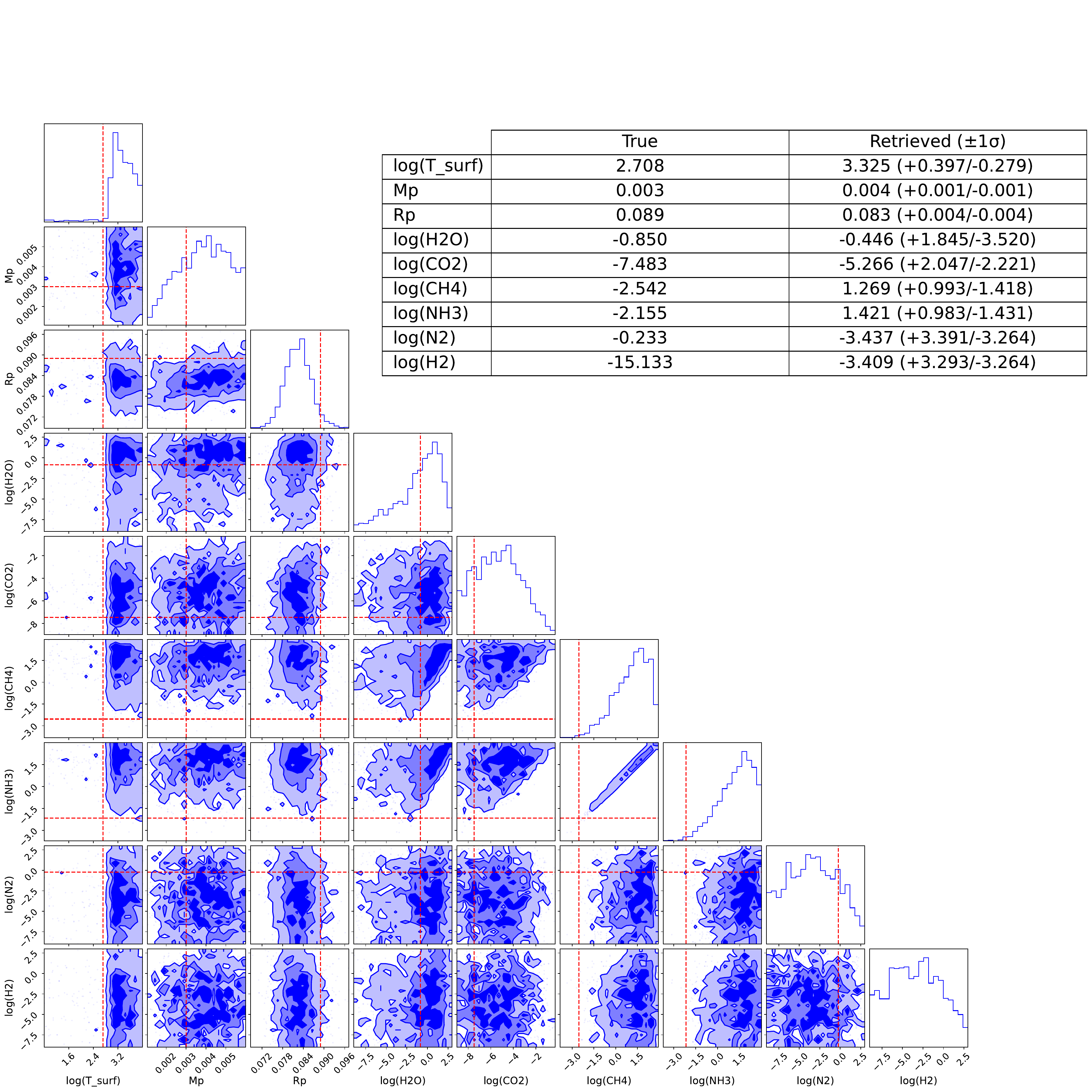}
    \caption{Same as in Fig.~\ref{fig:cornerIW+6} but for IW-2}
    \label{fig:cornerIW-2}
\end{figure*}

\begin{figure*}
    \centering
    \includegraphics[width=1\linewidth]{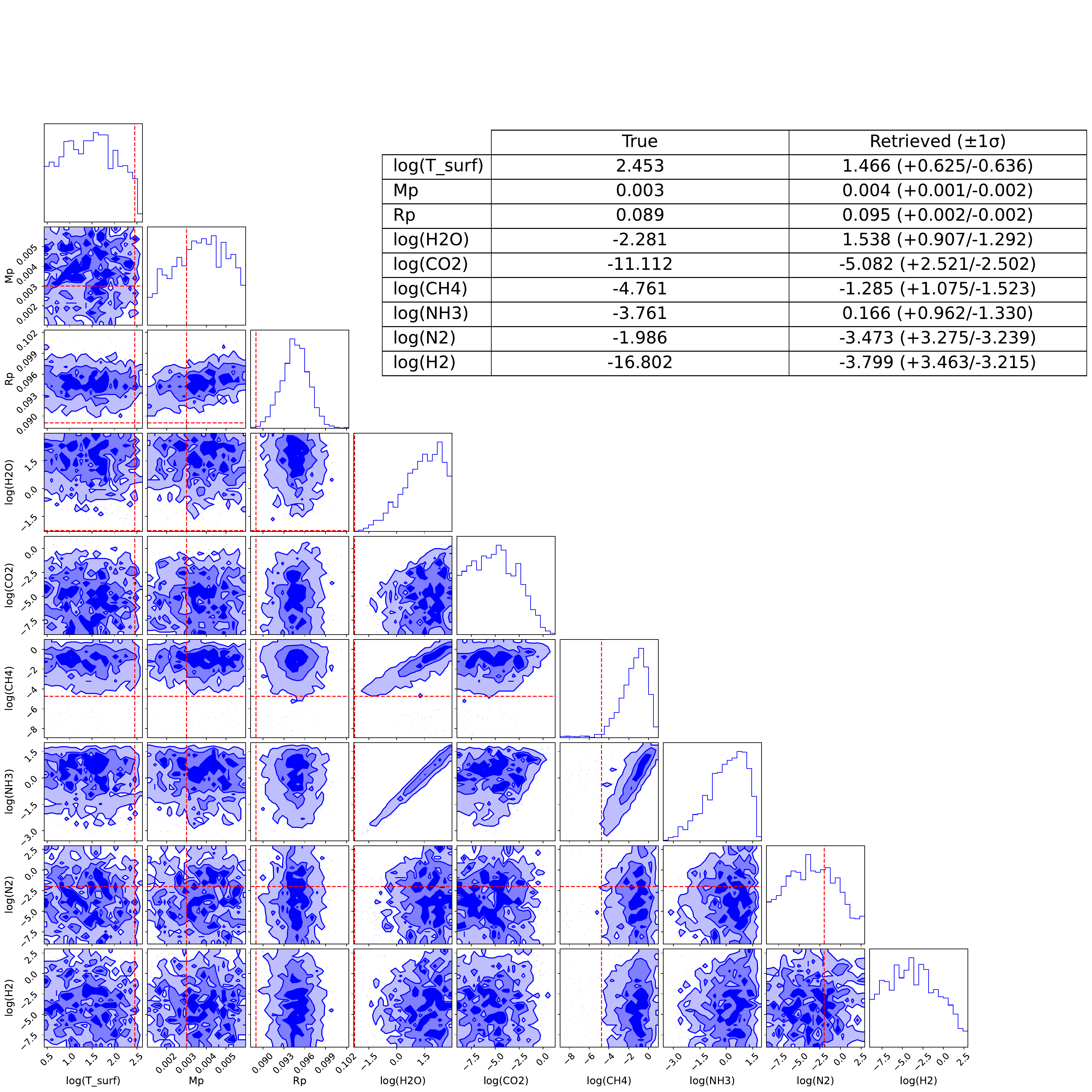}
    \caption{Same as in Fig.~\ref{fig:cornerIW+6} but for IW-4}
    \label{fig:cornerIW-4}
\end{figure*}

\begin{figure*}
    \centering
    \includegraphics[width=1\linewidth]{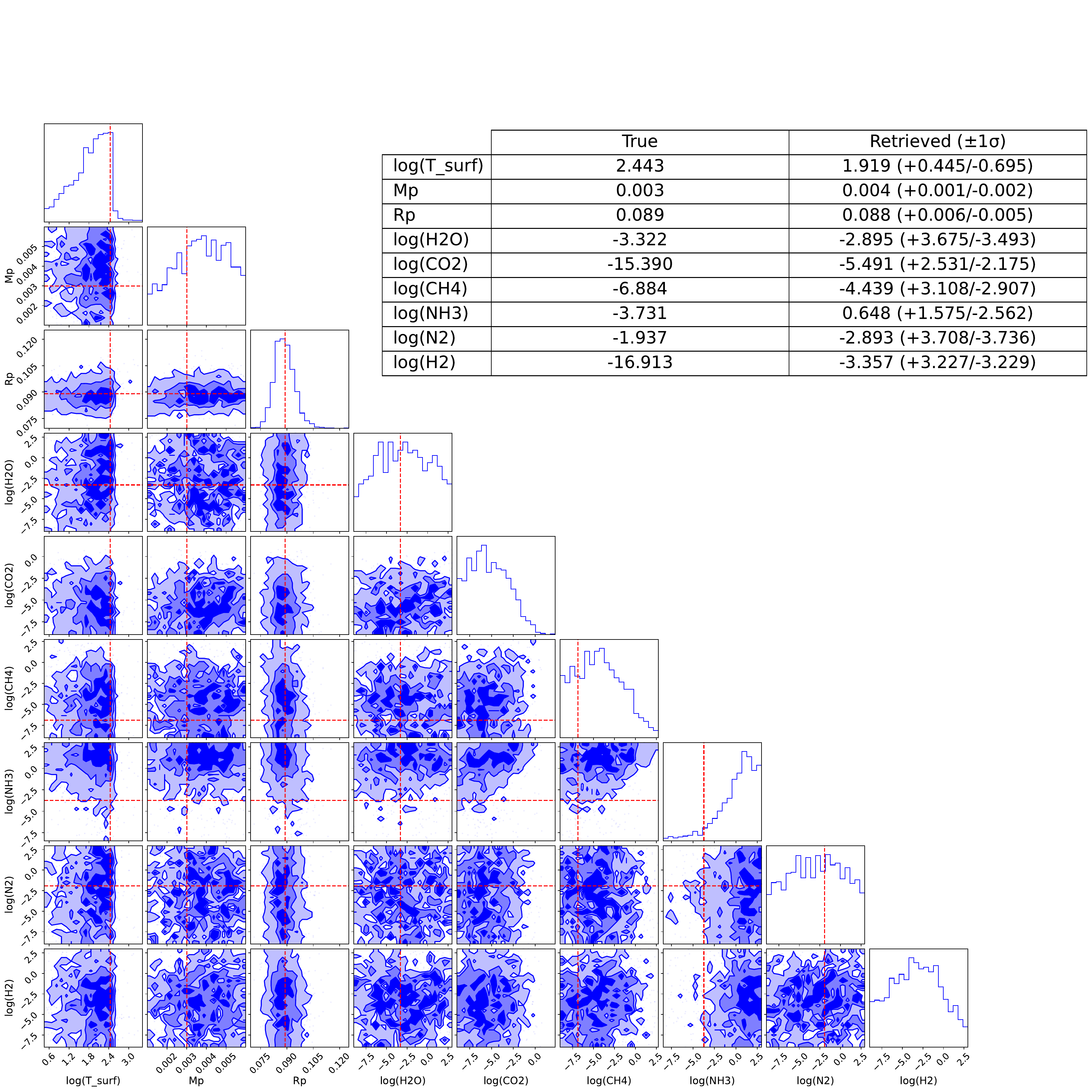}
    \caption{Same as in Fig.~\ref{fig:cornerIW+6} but for IW-6}
    \label{fig:cornerIW-6}
\end{figure*}

\hypersetup{
    allcolors=blue
}

\twocolumn

\section{Partition Coefficients and Solubility Data}
\label{sec:appendixC}

\renewcommand{\arraystretch}{0.7}
\begin{table*}
\centering
\caption{Solubility data for the six atmospheric species considered in this work, as used in the coupled mantle atmosphere degassing model of \citet{2025Icar..42916450B}. For H$_2$O, H$_2$, CO$_2$, and CH$_4$, the partial-pressure-dependent solubility in the melt follows the Henrian law of Eq.(6) of \citet{2025Icar..42916450B}; the underlying experimental references are listed in the last column. Nitrogen solubility is treated separately via the $f_{\mathrm{O}_2}$ dependent formulation of \citet{2003GeCoA..67.4123L}, Eqs.7-8 of \citet{2025Icar..42916450B}. NH$_3$ is not directly partitioned between melt and gas but forms as an atmospheric equilibrium product.}
\label{tab:solubility}
\begin{tabular}{lccl}
\hline
Molecule & $\alpha$ (ppm/Pa) & $\beta$ & References \\
\hline
H$_2$   & $2.572\times10^{-6}$ & $1.000$ & \cite{2003GeCoA..67.2427G}; \cite{2012hirsch} \\
H$_2$O  & $1.033$              & $1.747$ & \cite{1990CoMP..104..142S}; \cite{1995AmMin..80...94H}; \cite{1998CoMP..130..304M}; \\
        &                      &         & \cite{1999JPet...40.1497Y}; \cite{1999EPSL.168..201G}; \cite{2005JVGR..143..219L} \\
CO$_2$  & $1.937\times10^{-9}$ & $0.714$ & \cite{1975CoMP...53..227M}; \cite{1988EPSL...87..397S}; \cite{1991GeCoA..55.1587P}; \\
&&&                                        \cite{1993EPSL.119...27B}; \cite{1995JPet...36.1607D} \\
CH$_4$  & $9.937\times10^{-8}$ & $1.000$ & \cite{2013GeCoA.114...52A}; \cite{2019GeocL..12...12K} \\
N$_2$   & \multicolumn{2}{c}{---}        & \cite{2003GeCoA..67.4123L} \\
NH$_3$  & \multicolumn{2}{c}{---}        & Reaction R4 in \cite{2025Icar..42916450B} \\
\hline
\end{tabular}
\end{table*}

\end{document}